\documentclass[floatfix,10pt,superscriptaddress,twocolumn,showpacs,showkeys,aps,prb,final]{revtex4-1}
\usepackage{graphicx}
\usepackage{dcolumn}
\usepackage{bm}
\usepackage{ifpdf,braket}
\usepackage{epstopdf}

\begin{document}

\title{The linear Dirac spectrum and the Weyl states in the Drude-Sommerfeld topological model}
\author{Mauro M. Doria}
\affiliation{Instituto de F\'{\i}sica, Universidade Federal do Rio de Janeiro, 21941-972 Rio de Janeiro, Brazil}%
\affiliation{Instituto de F\'{\i}sica ``Gleg Wataghin'', Universidade Estadual de Campinas,\\ Unicamp 13083-970, Campinas, S\~ao Paulo, Brazil }
\email{mauromdoria@gmail.com}
\date{\today}

\begin{abstract}
A Drude-Sommerfeld topological model (DSTM) is proposed to describe Weyl fermions under residual collisions.
They are nearly free and dressed by their own weak magnetic field that breaks the reflection and time symmetries around a layer.
This weak magnetic field brings topological stability to the  states through a  non-trivial Chern-Simons number which is here calculated in the limit of a Dirac linear spectrum. The Weyl fermions display an energy gap and much above this gap the spectrum becomes  Dirac linear.
They are obtained from a Schroedinger like hamiltonian for particles with spin and magnetic energy which are momentum confined to a layer~\cite{doria17}.
The electrical and the thermal conductivities of the Weyl fermions as well as the corresponding Wiedemman-Franz law are derived in the framework of a constant relaxation time.
The Lorenz number coefficient acquires asymptotic value $6.5552$ times the bulk value of $\pi^2/3$.
The relaxation time is shown to be renormalized by the inverse of the square of the gap, and so, leads to a ballistic regime in the linear Dirac spectrum limit.
\end{abstract}

\maketitle
\section{introduction}
Two-dimensional materials display a wide range of electronic properties offering new venues for the development of nanoelectronic applications~\cite{miro14, nevalaita18}.
For this reason the study of the electronic properties of single two-dimensional layers, of surfaces and  of layered materials, are in the forefront of condensed matter physics research nowadays.
It has been known since long ago that surface states can have distinct properties from the bulk, transforming insulators into metals~\cite{tamm32, shockley39,hsieh08,xia09} and even into superconductors~\cite{bozovic14,uchihashi17,brun17}.
The extraordinary properties of graphene unleashed an intense search to identify  similar two-dimensional crystal lattices~\cite{geim07,wang15,young15}.
According to current theories the linear Dirac spectrum observed in graphene stems from its  hexagonal structure~\cite{wallace47,saito98}.  Electrons or holes near the six corners of the two-dimensional Brillouin zone act as  relativistic massless particles described by the Weyl equation. However much before the discovery of graphene, A.A. Abrikosov~\cite{abrikosov98,abrikosov99,abrikosov00,abrikosov03,zhang11} explained the linear magnetoresistance observed in nonstoichiometric silver chalcogenides~\cite{xu97}, and also in layered rare-earth diantimonides~\cite{budko98}, through a linear Dirac energy spectrum derived from a Weyl equation. While for graphene the Pauli matrices of the Weyl equation represent isospin, associated to the ``tight binding" hoping between the two intertwined lattices that form the hexagonal lattice~\cite{peter11}, for the chalcogenides and for the diantimonides A.A. Abrikosov just took for granted the Weyl equation without any assumption about the underlying lattice symmetry.\\

The linear Dirac spectrum has been observed in a variety of materials, among them a silicon layer~\cite{sadeddine2017}, the topological insulators~\cite{zhang09}, an organic conductor~\cite{hirata16}, the semimetals~\cite{li17,young15,hasan17}, the iron-based superconductors ~\cite{liu16,huynh11,terashima18}, the centrosymmetric superconductor {$\beta$-PdBi$_2$~\cite{sakano15}. The linear Dirac spectrum is present in many systems that display a linear magnetoresistance~\cite{lee02,he14,liang14,narayan15,xing16}.
In this paper a new scenario is considered for the onset of the Weyl states solely based on the breaking of the time and the reflection symmetry set by a layer~\cite{doria17}.
This scenario is fundamentally distinct from all the previous ones since it is not  based on the hopping between intertwined lattices nor on the crossing of energy bands.
Here the Pauli matrices of the Weyl equation  do represent the spin and from this scenario it emerges the remarkable fact that in the linear Dirac spectrum states are topologically protected  by their own magnetic field.
Hence the DSTM  consists of nearly free particles in a layer occupying Weyl states and  subjected to residual collisions, similarly to the original models proposed by Paul Drude in 1900 and extended with quantum concepts by Arnold Sommerfeld in 1927.\\

The breaking of the space-time symmetries around the layer is the key ingredient that renders the present proposal distinct from the standard Drude-Sommerfeld model. The onset of the  magnetic field produced by the particles  breaks such symmetries, and this field  cannot be dismissed  no matter how small it is, since it has topological consequences.
In the traditional Drude-Sommerfeld scenario the particles move freely  between the residual collisions and the same holds for the present DSTM. Nevertheless the magnetic field is dismissed in the traditional view and here is included. Its three-dimensional configuration is explicitly obtained, and shown to lead to topologically protected states independently of its strength. The Chern-Symons (topological) index associated to this local magnetic field is analytically obtained and found to be non-trivial. The magnetic field streamlines form loops that pierce the layer twice and cannot be broken unless by a strong collision regime.
However the DSTM is in a weak collision regime because the collision time becomes increasingly large, as shown here. Hence a consistent picture emerges of topologically stable free particles in a Drude-Sommerfeld scenario.\\

The collision regime, characterized by the collision time $\tau$,  yields an  electrical ($\sigma$) and a thermal ($\kappa$) conductivity for the DSTM, which are obtained in this paper. The existence of the ballistic regime is a consequence of the shape of the Fermi surface, which has an inner and an outer surface as the dispersion relation has a mexican hat like shape. Recently such Fermi surfaces have been investigated in novel  electronic systems~\cite{stauber07,wickramaratne15,jo17}. In the limit of the linear Dirac spectrum the Fermi surface retains its unusual properties because it becomes  a disk removed of its center.
Here the Fermi surface is taken smaller than the size of the Brillouin zone for the sake of simplicity. The onset of Fermi arcs is possible from the present approach, in case the Fermi surface is larger than the size of the Brillouin zone, as shown in Ref.~\onlinecite{doria17}. For simplicity $\tau$ does not include any corrections due to phonon, back scattering, short and long range scattering potentials~\cite{peres07}, and other sources, such as the Bloch-Gr\"uneisen temperature whose effects in the resistivity have been studied for the  two-dimensional electron gas~\cite{stormer90} and for graphene~\cite{efetov10}. Possible effects on the scattering among particles caused by the local magnetic field are also not included. A more elaborate study of the resistivity including these features will be seen elsewhere. The main goal here is simply to show that the electrical and thermal conductivities in the linear Dirac spectrum limit fall in a ballistic regime regardless of the inclusion  of such effects. For this purpose it suffices the simplest treatment of a constant and isotropic $\tau$. The Wiedemann-Franz law is obtained as the ratio $\sigma/\kappa$ is independent of the collision time. The coefficient of the Lorenz ratio ($(\sigma/T \kappa)/(k_B/q)^2)$) is found to be distinct from the standard bulk value of $\pi^2/3$.\\

\begin{figure}[ht]
\center
\includegraphics[width=\columnwidth]{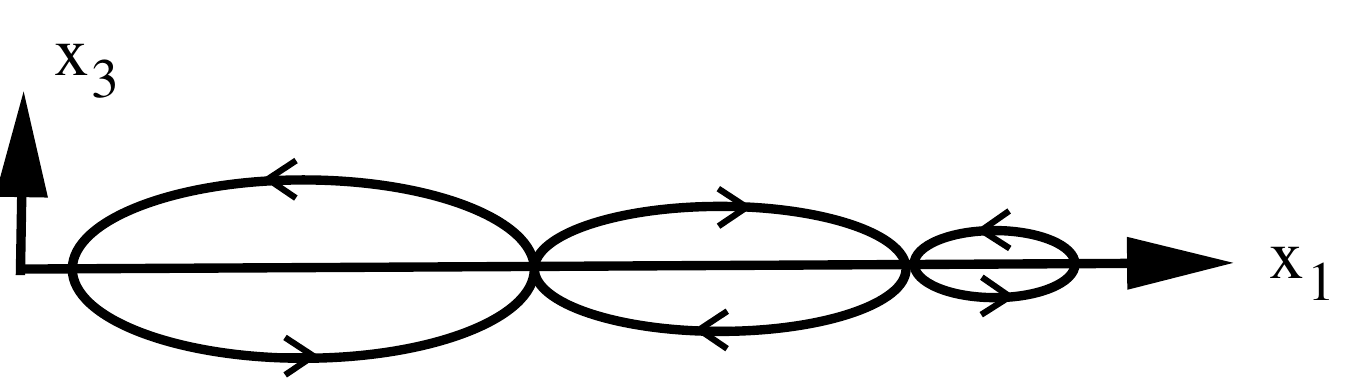}
\caption{A pictorial view of the closed magnetic field stream lines that pierce the layer twice. These stream lines dress the linear Dirac spectrum particle in the layer. The magnetic field obtained from the theory (Eq.(\ref{avgh})) is shown in Figs.~\ref{figfield} and ~\ref{figcont}.}
\label{figstreamline}
\end{figure}

This work is organized as follows.
In Sec.~\ref{toposec} the hamiltonian of the Drude-Sommerfeld and its topological properties are discussed.
In Sec.~\ref{weylsec} the breaking of the space-time symmetries and the onset of the magnetic field is considered.
In Sec.~\ref{electhersec} the electrical and thermal conductivities of the Weyl state are obtained.
In Subsec.~\ref{dimensubsec} the parameters and dimensionless units of the present theory are given.
In Subsec.~\ref{numsubsec} the number of particles operators are discussed.
In Subsec.~\ref{elecsubsec} the electrical conductivity is derived based on the Boltzmann-BGK equation, and similarly,
in Subsec.~\ref{thersubsec} the thermal conductivity is obtained.
In Subsec.~\ref{lindirsubsec} the conductivities and the number of particles in the linear Dirac spectrum limit are obtained.
In Subsec.~\ref{lowhigsubsec} the low and high temperature limits of the conductivities are studied and the chemical potential obtained as a function of the temperature.
In Subsec.~\ref{wiedsubsec} the Wiedemann-Franz law at the low and high temperature limits are derived.
In Sec.~\ref{magsec} the  magnetic field effects are included and a general discussion about the attractive magnetic energy is given.
In Subsec.~\ref{estimsubsec} the  magnetic field and  the magnetic energy are estimated and consequently the free particle picture confirmed.
In the Conclusion, Sec.~\ref{conclusion}, the Drude-Sommerfeld scenario of free particles with residual interactions is summarized.
The appendix contains three sections.
In App.~\ref{appa} the integration in energy for the two possible momentum branches is studied.
In App.~\ref{appb} some useful integrals are obtained.
In App.~\ref{appc} the  Chern-Symons index is analytically determined.

\section{Topological properties of the Drude-Sommerfeld model}\label{toposec}
The hamiltonian of the Drude-Sommerfeld framework  provides the thermally accessible energy levels $E$ that are accessed through the residual scattering  near to the Fermi level. The residual scattering can be formally included through the non-equilibrium Boltzmann-BGK equation~\cite{ashcroft76,coelho17}.
The occupation number of an electronic state $\vert \chi \rangle$, $H\vert \chi \rangle=E\vert \chi \rangle$, is
$f_0(\vec k) = 1/\left \{ \exp{[\beta(E-\mu )]}+1\right \}$,  $\beta \equiv 1/k_B T $, for a temperature $T$ and  $\mu$ is the chemical potential.
The two new key ingredients of the DSTM with respect to the standard Drude-Sommerfeld theory are the inclusion of the electronic spin and of the residual local magnetic field created by the existing currents that enter through Amp\`ere's law. Hence the hamiltonian is given by,
\begin{eqnarray} \label{dseq}
&& H=\int d^3 \vec r \left [\frac{1}{2m}\vert \vec P\Psi \vert ^2+
\frac{1}{8\pi} \vec h(\Psi)^2 \right ] ,\quad  \Psi =
\left(
\begin{array}{c} \psi_{\uparrow} \\ \psi_{\downarrow} \end{array} \right),\\
&& \vec \nabla \times\vec h(\Psi) = \frac{4\pi}{c}\vec J(\Psi), \;
\vec J = \frac{q}{2m}\left (\Psi^* \vec P \Psi \, + \, \mbox{c.c.} \right ),
\end{eqnarray}
where $\vec P = (\hbar/i)\vec \nabla - (q/c)\vec A$ and $\vec h =\vec \nabla \times \vec  A$ is the local magnetic field.
The current $\vec J(\Psi) $ created by the particles, described by $\Psi$, is the source of the local magnetic field $\vec h(\Psi)$.\\

The most remarkable  property of the above theory is its ability to describe Weyl states when applied to a layer. The Weyl state satisfies the equation below.
\begin{eqnarray}\label{weyl}
v_0 \vec \sigma \cdot \vec P \Psi = E_0 \Psi,
\end{eqnarray}
where the Pauli matrices $\vec \sigma$ represent spin, whereas the linear Dirac energy spectrum is given by,
\begin{eqnarray} \label{dirac}
E(k) = v_0 \hbar \vert \vec k \vert, \; \vec k = k_1 \hat x_1 + k_2 \hat x_2.
\end{eqnarray}
Strictly speaking the Weyl state only has a  Dirac linear spectrum for $E_0=0$, hereafter referred as a {\it zero helicity state} (ZHS)~\cite{doria17} because $\vec \sigma \cdot \vec P $ is proportional to the projection of the total angular momentum, $\vec r \times \vec P + \hbar \vec \sigma/2$, along momentum $\vec P$.
Interestingly the spin-locking condition is automatically satisfied by the ZHS at a local level since it means that the spin is locked at right angles to the momentum~\cite{li14}, and so, has zero helicity.
This property has been observed in the topological insulators, which  exhibit metallic surface states with a linear Dirac  spectrum.
This state has also been called as the spin helical Dirac transport regime~\cite{hsieh09}.\\

The ZHS features the remarkable property that the three-dimensional  magnetic field distribution around the layer can be exactly known since Amp\`ere's law is exactly. The ZHS satisfies  the following equations,
\begin{eqnarray}
&& \vec \sigma \cdot \vec P \Psi  = 0, \label{foe1}\\
&& \vec h  = -4\pi \mu_B\Psi^{\dag}\vec \sigma \Psi, \label{foe2}
\end{eqnarray}
to be later explained, where we have introduced the Bohr's magneton, $\mu_B=q\hbar/2 m c$.
Notice that according to the above equation $\vec h$ and $\vec \sigma$ must have the same intrinsic symmetry properties which means that they are pseudo vectors.
The magnetic field is divergenceless, $\vec \nabla \cdot \vec h =0$, and so the magnetic field stream lines are closed, as depicted in Fig.~\ref{figstreamline}.
Assume  periodicity within the layer by a square unit cell with side $L$, described by the planar coordinates $(x_1,x_2)$, such that at each of its points a unit vector is defined by the direction of this magnetic field associated to the state $\vert \chi \rangle$: $\hat h = \langle \vec h\rangle/ \vert \langle \vec h \rangle \vert$, where
$\langle  \vec h \rangle = \langle \chi| \vec h |\chi \rangle= -4\pi \mu_B \langle \chi| \Psi^{\dag}\vec \sigma \Psi|\chi \rangle $. Therefore a mapping from a torus into a sphere has been established and its Chern-Symons index is given by,
\begin{eqnarray}\label{topo}
Q= \frac{1}{4\pi}\int_{x_3=0^+} \big (\frac{\partial \hat
h}{\partial x_1} \times \frac{\partial \hat h}{\partial x_2}
\big)\cdot \hat h \; d^2x.
\end{eqnarray}
In this paper we analytically obtain $Q$ for single particle states and show that they are topologically non-trivial states ($Q \ne 0$). Hence the scenario is of  particles with a Dirac linear spectrum  dressed by their own magnetic field and  protected to decay into lower energy states. These dressed particles are skyrmion like states similar to those found in other condensed matter systems~\cite{dupe14,bazeia16,tsesses18}.

\section{The Weyl State and the breaking of the space-time symmetries} \label{weylsec}
In this section the symmetries of the Weyl state are discussed on light of the invariance of the kinetic energy.
A time-reversal operation flips the direction of the local field,  $\vec h \rightarrow -\vec h$, and so reverts the direction of the closed stream line. A reflection operation, $x_3 \rightarrow -x_3$, flips the top and the bottom half spaces, defined with respect to the layer, and so creates a dressed state with its magnetic field  mirror reflected. If the stream line is symmetric with respect to the layer, as here, the reflection operation has the same effect of the time reversal symmetry on the closed stream line.  In conclusion the local magnetic field is only possible if either the time or the reflection symmetries are independently broken since their product is conserved. Indeed a detailed analysis of the Eqs.(\ref{foe1}) and (\ref{foe2}) shows that they  violate the time and the reflection  symmetries but not the combined operations, which  leave the closed stream line invariant.
If the set $\Psi$, $\vec h$ is a solution, so is the set, $\Psi^{\prime}= \sigma_2\Psi^{*}$, $\vec h^{\prime}=-\vec h$.
It must be stressed at this point the inadequacy of the  Weyl equation, given by  Eq.(\ref{weyl}),  to describe electronic states as a hamiltonian. The helicity operator $\vec \sigma \cdot \vec P $ is a {\it pseudo scalar} and so is  the parameter  $E_0$. Consequently $E_0$ cannot describe energy, which is a scalar quantity. The symmetry properties of the helicity operator are briefly reviewed for the sake of completeness. Firstly notice that reflection in a layer and parity symmetries are equivalent.
Reflection is $(x_1,x_2,x_3) \rightarrow (x_1,x_2,-x_3)$ whereas  parity is $(x_1,x_2,x_3) \rightarrow (-x_1,-x_2,-x_3)$.
However they are equivalent since $(x_1,x_2,x_3) \rightarrow (-x_1,-x_2,x_3)$ is a pure rotation.
A parity transformation ($\vec r \rightarrow -\vec r $) applies to momentum $\vec P$ and gives
$\vec P \rightarrow -\vec P$, and leaves invariant the spatial angular momentum, $\vec r \times \vec P$ and so, also spin $\hbar \vec \sigma/2$, thus
$\vec \sigma \rightarrow + \vec \sigma$. Therefore the  helicity operator flips sign under the parity operation, $\vec \sigma \cdot \vec P \rightarrow -\vec \sigma \cdot \vec P $. In conclusion the Weyl equation (Eq.(\ref{weyl})) must be reinterpreted since it is not an eigenvalue problem. The pseudo scalar nature of $E_0$ is made explicit by introducing the more appropriate parameter $\theta$, shown to act as  a wavenumber cutoff.
\begin{eqnarray}
E_0\equiv v_0 \hbar \theta \frac{x_3}{\vert x_3\vert }
\end{eqnarray}
Prior to the interpretation of Eq.(\ref{weyl}), which is a central matter to this paper, its solution is given below.
\begin{eqnarray}\label{psiweyl}
&& \Psi(\vec{x}) = \frac{1}{\sqrt{V}} \sum_{\vec k,\; k > \theta} c_{\vec k} \; e^{i\vec{k}\cdot\vec{x}} e^{-l(k) \vert x_3 \vert  }
\left( \begin{array}{c} 1 \\ \frac{x_{3}}{\vert x_{3}\vert }\frac{k_{+}}{k} \left (\frac{\theta-i l(k)}{k}\right ) \end{array} \right), \nonumber\\
&& \mbox{where} \quad l(k) \equiv \sqrt{k^2-\theta^2},
\end{eqnarray}
$k \equiv \vert \vec k \vert = \sqrt{k_{+}k_{-}}$,  and $k_{\pm}\equiv k_1\pm i k_2$. Although the $l(k)$ imaginary solutions exist they are excluded because they describe propagating waves scattered by the layer. The solutions confined to the layer are those such that $l(k)> 0$, which render momentum confinement to the layer but not position confinement since the state exponential decays away from it. The volume $V=A L_3$ is the unit cell area ($A \equiv L^2$) times $L_3$, which is an arbitrary length perpendicular to the layer that defines the velocity parameter $v_0$. In conclusion this is a three-dimensional description of a state confined to a two-dimensional layer. The coefficient $c_{\vec k}$ is interpreted as the destruction operator of a Weyl fermion.\\

The ZHS, obtained in the limit $\theta \rightarrow 0$, is a state localized in the layer under exclusion of the zero momentum state, $k \ne 0$, since $l(k)=\vert \vec k \vert>0$. All the non-zero states  break rotational invariance since $k_3=\pm i\sqrt{k_1^2+k_2^2}$. The square of Eq.(\ref{foe1}), in the absence of a vector potential, gives that $\vec P^2\Psi=0$ and for a plane wave it follows that $k_1^2+k_2^2+k_3^2=0$. Thus  the only possible propagating solution is $k_1=k_2=k_3=0$. The ZHS is given by,
\begin{equation}\label{psizhs}
\Psi(\vec{x}) = \frac{1}{\sqrt{V}} \sum_{\vec k,\; k >0} c_{\vec k} \; e^{i\vec{k}\cdot\vec{x}} e^{-k \vert x_3 \vert  }
\left( \begin{array}{c} 1 \\  -i\frac{x_{3}}{\vert x_{3}\vert }\frac{k_{+}}{k} \end{array} \right).
\end{equation}
One can think of this state as a combination of a symmetric and of an anti-symmetric state across the layer, given by
$\left( \begin{array}{c} 1 \\  0\end{array} \right)$  and $\left( \begin{array}{c}  0 \\  \frac{x_{3}}{\vert x_{3}\vert }\ \end{array} \right)$, respectively, although the present approach does not take into account the thickness of the layer.
Interestingly a finite thickness film can present a degeneracy in energy between the symmetric and anti-symmetric states~\cite{doria16}.\\

It is straightforward to show that the local magnetic field consists of closed stream lines that pierce the layer twice.
Just notice that the magnetic field is obtained from Eq.(\ref{foe2}) and that according to Eq.(\ref{psizhs}) it holds that $\Psi^{\dag} \sigma_1 \Psi \sim x_3/\vert x_3 \vert$, $\Psi^{\dag} \sigma_2 \Psi \sim x_3/\vert x_3 \vert$ whereas $\Psi^{\dag} \sigma_3 \Psi$ does not depend on $x_3/\vert x_3 \vert$. Therefore the field components $h_1$ and $h_2$ flip sign from one side of the layer to the other while $h_3$ does not. The only way to fulfil these requirements is through a stream line that crosses the layer twice.\\

Next this symmetry breaking state is fitted into the kinetic description of the DSTM, which is invariant under such space-time transformations.
The key ingredient is a decomposition of the kinetic energy as the sum of the Weyl equation squared plus the Rashba interaction added to an interaction of the magnetic moment with the local field~\cite{doria17}. There it becomes clear that the symmetry breaking Weyl state does not upsets the symmetries of the underlying rotationally symmetric hamiltonian of Eq.(\ref{dseq}).
\begin{eqnarray}\label{kin01}
K = \int d^3x \; \frac{1}{2m}|\vec{P}\,\Psi|^2, \,
\end{eqnarray}
can be expressed as,
\begin{eqnarray} \label{kin02}
&& K = \int d^3x \; \left \{\frac{1}{2m}\vert\vec{\sigma}\cdot\vec P\Psi \vert^2-  \right . \\
&&  \left . \frac{\hbar}{4m}\vec \nabla \cdot \left[\Psi^\dag\left(\vec \sigma \times \vec P\right)\Psi+c.c. \right]
 +\frac{\hbar q }{2 m c}\vec{h} \cdot\left(\Psi^\dag\vec\sigma\Psi\right)\right \}. \nonumber
\end{eqnarray}
The three term  decomposition keeps the  rotational invariance of the kinetic energy. Notice that the Rashba  interaction  only exists in the two-dimensional layer. It is given by the surface term of the above kinetic energy, whereas the other two terms are three-dimensional. The choice of a Weyl state that satisfies Eq.(\ref{weyl}) does not upset this invariance although it implies the breaking of the time reversal and the reflection symmetries.
Thus the state has a lower symmetry than the theory itself, a situation commonly seen in physics in the context of the Goldstone theorem~\cite{hoinka17}. The three term decomposition of the kinetic energy is general and can be applied in other situations such as in the Ginzburg-Landau approach to layered superconductors~\cite{alfredo13,doria14,alfredo15,edinardo16,edinardo17} where the pseudogap can be interpreted as a topological state~\cite{cariglia14}.\\

The kinetic energy under the assumption of a Weyl state becomes,
\begin{eqnarray} \label{kinweyl}
&& K = \int d^3x \; \left \{\frac{\hbar^2\theta^2}{2m}\Psi^\dag\Psi + \right . \nonumber \\
&& \left . \frac{\hbar^2}{4m}\nabla^2 \left (\Psi^\dag \Psi \right )+ \frac{\hbar q }{2 m c}\vec{h} \cdot\left(\Psi^\dag\vec\sigma\Psi\right) \right \}.
\end{eqnarray}
Notice that the $\theta^2$ term is a consequence of the  Weyl state and therefore the Weyl equation {\it is not} directly responsible for the linear Dirac spectrum. It is indirectly responsible though since the linear Dirac spectrum stems from the Rashba term, as shown in Ref.~\onlinecite{doria17}, under the condition of a Weyl state.
In summary the breaking of the space-time symmetries allows for the onset of the local magnetic field that dresses the particle. As shown in this paper this field is  weak and gives topological stability to the particles.

\begin{figure}[ht]
\center
\includegraphics[width=\columnwidth]{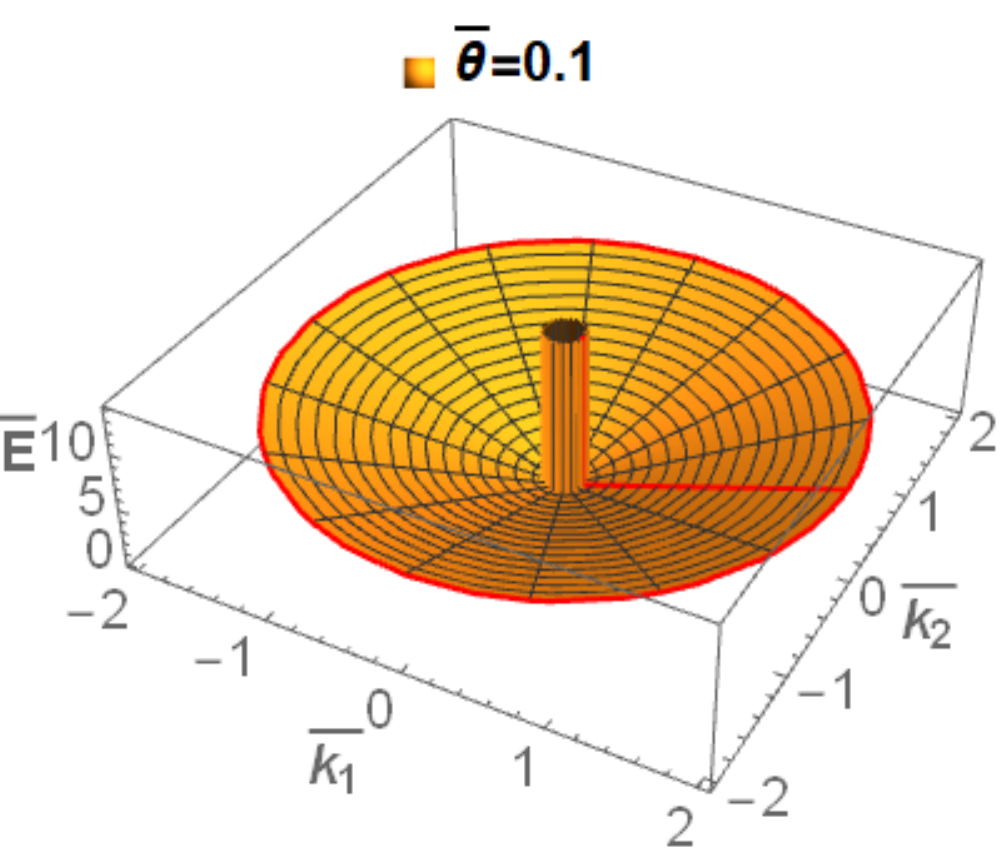}
\caption{A three dimensional view of the Weyl state dispersion relation, given by Eq.(\ref{bare}), versus the two-dimensional wavenumber, $(\bar k_1, \bar k_2)$, is shown here for a particular value of the wavenumber cutoff, $\bar \theta$.}
\label{figthreed}
\end{figure}
\begin{figure}[ht]
\center
\includegraphics[width=\columnwidth]{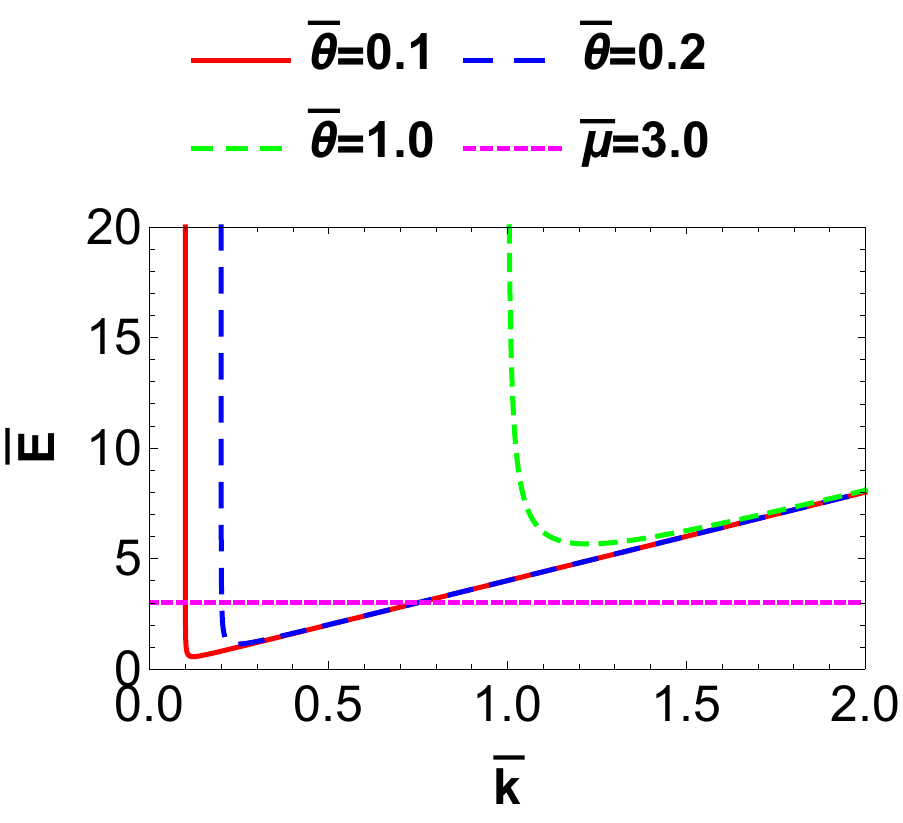}
\caption{The Weyl state dispersion relation, given by Eq.(\ref{bare}), versus the wavenumber, $\bar k = \sqrt{\bar k_1^2+ \bar k_2^2}$, is shown for several values of the wavenumber cutoff, $\bar \theta$, and for a fixed value of the chemical potential, $\bar \mu$.}
\label{figdisper}
\end{figure}
\section{The electrical and thermal conductivities of the Weyl state}\label{electhersec}
In this section the local magnetic field that dresses the particles is taken weak enough that the electrical and the thermal conductivities are obtained without its presence. The weakness of the local magnetic field is discussed in Sec.~\ref{estimsubsec}. For the purpose of the derivation of the conductivities it suffices to consider the energy levels $E$ that stem from the hamiltonian removed of its magnetic field content.
The collision time $\tau$ is assumed to be momentum and temperature independent, which oversimplifies the model. For this reason the ratio between the electrical and thermal conductivities (the Wiedemann-Franz law), is studied since it does not depend on $\tau$.\\
The main result found here is the discovery of a ballistic regime in the linear Dirac spectrum limit ($\theta \rightarrow 0$). This is equivalent to say that the collision time is renormalized and becomes equal to $\tau/\theta^2$.
This brings support to the scenario of unbroken magnetic field stream lines under collisions as those become scarce.
\begin{eqnarray}
H=K \approx \int d^3x \;\left \{ \frac{\hbar^2\theta^2}{2m}\Psi^\dag\Psi +
\frac{\hbar^2}{4m}\nabla^2 \left (\Psi^\dag \Psi \right)\right \}
\end{eqnarray}
Introducing the Weyl state of Eq.(\ref{weyl}) one obtains that in momentum space the above kinetic energy becomes,
\begin{eqnarray}
&& K = \sum_{\vec k,\; k \ge \theta}E(k){c_{\vec k}}^\dag c_{\vec k},  \\
&& E(k) \equiv \frac{\hbar^2}{2mL_3}\left [2 \frac{\theta^2}{\sqrt{k^2-\theta^2}}
+4\sqrt{k^2-\theta^2}\right ] \label{disprel}
\end{eqnarray}
There is an energy gap, $\Delta$, in this dispersion relation,
\begin{eqnarray}
\Delta \equiv E(k_{min}) = 2 \sqrt{2} \frac{\hbar^2}{mL_3}\theta, \quad k_{min}=\frac{3}{2}\theta.
\label{mu}
\end{eqnarray}
Interestingly this dispersion relation leads to a Fermi surface in the shape of a ring, a subject that  has been discussed both theoretically and experimentally~\cite{stauber07,wickramaratne15,jo17}. The study of this dispersion relation is done using the dimensionless units described below.

\subsection{Parameters and dimensionless units}\label{dimensubsec}
The present theory contains two independent parameters, namely, the particle density in the layer and the in-plane velocity,
\begin{eqnarray}
&& n \equiv \frac{N}{L^2}, \, \mbox{and}\\
&& v_0 \equiv \frac{2\hbar}{m L_3},
\end{eqnarray}
respectively, the latter defined by the out of plane length $L_3$.
The linear  Dirac spectrum is retrieved  in the limit $k >> \theta$ since $E(\vec k) \approx v_0 \, \hbar \,k$. In case that $\theta \rightarrow 0$ this holds for all $k$ since $E(\vec k) \rightarrow v_0 \, \hbar \,k$. The Fermi wavenumber in the linear Dirac spectrum is obtained by filling particles up to the Fermi surface, which is a disk  with radius $k_F$ removed of its center $k = 0$: $N=\sum_{\vec k}= (L/2\pi)^2\int d^2 \vec k = (L/2\pi)^2 \pi k_F^2$, and so, $k_F=\sqrt{4\pi n}$.
The $\theta=0$ Fermi surface is  $E_F=\hbar v_0 k_F$, that can also be expressed as  $E_F = 4e_0 L_3 k_F$ or $E_F = e_0 \sqrt{n/n_0}$ where,
\begin{eqnarray}
e_0 = \frac{\hbar^2}{2mL_3^2},
\end{eqnarray}
is a unit of energy, and
\begin{eqnarray}
n_0 =\frac{1}{64\pi L_3^2}.
\end{eqnarray}
is a unit of density.
The units of electrical conductivity and of thermal conductivity are given by
\begin{eqnarray}
\sigma_0 = \frac{q^2}{m L_3^3}\tau,
\end{eqnarray}
and
\begin{eqnarray}
\kappa_0 = \sigma_0\left( \frac{k_B}{q}\right)^2 T,
\end{eqnarray}
respectively. Therefore the  {\bf dimensionless units} are defined by $e_0$ (energy), $L_3$ (length),
$n_0$ (density), $\sigma_0$ (electrical conductivity), and $\kappa_0$ (thermal conductivity).
Dimensionless quantities carry a ``bar" over the respective symbol.
For instance, the gap becomes $\Delta \equiv  e_0 \bar \Delta$,  $\bar \Delta \equiv \sqrt{2}\, 4\bar \theta$, the dimensionless chemical potential is defined by  $ \mu \equiv e_0 \bar \mu$, and other dimensionless quantities follow similarly, such as the reduced temperature, $\bar \beta \equiv e_0 \beta$, and several wavenumbers, namely,
$\bar k \equiv L_3 k$, $\bar \theta = L_3 \theta$, and $\bar l = L_3 l$. Notice that $\bar \beta \bar \mu = \beta \mu$. The dimensionless density $\bar n$ is defined through $n = n_0 \bar n$.
The Weyl state dispersion relation becomes $E(k,\theta) = e_0 \bar E(\bar k, \bar \theta)$,
\begin{eqnarray}\label{bare}
\bar E =\left [2\frac{\bar \theta^2}{\bar l}+ 4 \bar l \right ], \quad \bar l \equiv \sqrt{\bar k^2-\bar \theta^2}.
\end{eqnarray}
Therefore in the linear Dirac spectrum the Fermi surface becomes $\bar E_F = \sqrt{\bar n}$ since $E_F=e_0 \bar E_F$.\\

Figs.~\ref{figthreed} and ~\ref{figdisper} show the above dispersion relation in dimensionless units. The mexican hat view of Fig.~\ref{figthreed} shows a three-dimensional dispersion relation for $\bar \theta=0.1$. Notice the center of the hat is excluded since it can never be reached under a finite chemical potential. The projection of this plot into two-dimensions renders a ring like dispersion relation with an inner and an outer radii. A sectional view of the mexican hat is given by  Fig.~\ref{figdisper} and shows the dispersion relation for several $\bar \theta$ and a single $\bar \mu$. The $\bar \theta=0.1$ and $\bar \theta=0.2$ states are of conductors since the chemical potential lies above the minimum whereas the $\bar \theta=1.0$ state is of an insulator since there are no available states for conduction.
In the latter case the chemical potential falls below the minimum of the dispersion relation.
The linear Dirac spectrum valid for $\bar k \gg \bar \theta$ is clearly seen in both Figs.~\ref{figthreed} and ~\ref{figdisper}.\\

The  Fermi surface is defined at the intersect of the chemical potential with  the energy dispersion relation, $\bar E_F \equiv \bar \mu=\bar E$. This happens  at the inner and outer Fermi wavenumbers given below~\cite{doria17} .
\begin{eqnarray}
&& \bar k_{aF}^2 = \frac{3}{2}\bar \theta^2 + \frac{\bar \mu^2}{32} \left [\bar \delta^2 + \bar \delta \right ], \label{kaf}\\
&& \bar k_{bF}^2 = \frac{3}{2}\bar \theta^2 + \frac{\bar \mu^2}{32} \left [\bar \delta^2 - \bar \delta \right ], \label{kbf} \\
&& \bar \delta \equiv \sqrt{1-\left(\frac{\bar \Delta}{\bar \mu} \right)^2},\label{delta}
\end{eqnarray}
Therefore the Weyl state has two Fermi surfaces yielding a ring (mexican hat) shape.
The chemical potential touches the bottom of the band at $\bar \mu = \bar \Delta$, then $\bar \delta=0$, and both wavenumbers collapse into a single one, $\bar k_{aF}=\bar k_{bF}=\bar k_{min}=\sqrt{3/2}\bar \theta$. The Fermi surface is reduced to a circle in this case. Away from this minimum the Fermi wavenumbers fall below and above the critical value, namely, $\bar k_{bF} \le \bar k_{min}$ and $\bar k_{aF} \ge \bar k_{min}$, respectively.
Interestingly the inner Fermi surface avoids the evasion of particles from the layer.
The Weyl state of Eq.(\ref{weyl}) decays exponentially away from the layer, $\Psi \propto \exp{\left (- \sqrt{k^2-\theta^2}\vert x_3 \vert \right)}$, and so, for $k=\theta$ the particle becomes delocalized from the layer.
However this never takes place because it always holds that $\bar k_{bF} \ge \bar \theta$. In the limit $\bar \theta \rightarrow 0$, and for  $\bar \mu$ fixed,  $\bar k_{aF} \rightarrow \bar \mu/4$  and $\bar k_{bF} \rightarrow \bar  \theta \rightarrow 0$. This show that the Fermi surface becomes a disk removed of its center in the linear Dirac spectrum limit.\\

The dimensionless electrical and thermal conductivities are defined through $\sigma = \sigma_0 \bar \sigma$ and $\kappa = \kappa_0 \bar \kappa$, respectively such that,
\begin{eqnarray}\label{kapsig0}
\frac{\kappa}{T \sigma} = \frac{\bar \kappa}{\bar \sigma }\left ( \frac{k_B}{q}\right)^2
\end{eqnarray}
Therefore the Lorenz coefficient of  the Wiedemann-Franz law is just the ratio between the dimensionless thermal and electrical conductivities, $\bar \kappa /\bar \sigma$, known to be equal to $\pi^2/3$ for the bulk.

\subsection{The number of particles}\label{numsubsec}
The number of particles in the original hamiltonian of Eq.(\ref{dseq}) can be adjusted by taking $H-\tilde \mu \tilde N$  where $\tilde N = \int d^3x \; \Psi^\dag \Psi$.
Introducing the $\Psi$ of a Weyl state, as given by Eq.(\ref{psiweyl}), this number of particle operator becomes,
\begin{eqnarray}
\tilde N = \frac{1}{L_3} \sum_{\vec k,\; k > \theta}\frac{2}{\sqrt{k^2-\theta^2}}\,c_{\vec k}^{\dag} c_{\vec k}.
\end{eqnarray}
However there is another number of particles operator, given by,
\begin{eqnarray}
N=\sum_{\vec k,\; k > \theta} \, c_{\vec k}^{\dag} c_{\vec k}.
\end{eqnarray}
We interpret that $\tilde N$ and $N$ describe the number of original particles and of  Weyl quasi-particles, respectively, both expressed in terms of quasi-particle operators. Similarly $\tilde \mu$ and $\mu$ are their respective chemical potentials. We assume here for simplicity that the total number of particles is equal to that of quasi-particles. At temperature $T$ the Weyl states are filled according to  $N = \sum_{\vec k, \, k>\theta} f_0(\vec k,T)$  such that in terms of dimensionless variables one gets that,
\begin{eqnarray}
&&n=\frac{1}{4\pi L_3^2} \int_0^{\infty} d \bar l \, (2\bar l) f_0(\bar l) \rightarrow   \\
&& \bar n =  16 \int_0^{\infty} d \bar l \, \frac{d\bar l^2}{d \bar l} f_0(\bar l).
\end{eqnarray}
Integration by parts plus using that $df_0(\bar l)/d \bar l= (\partial f_0/\partial \bar E) \partial \bar E/\partial \bar l$, gives that,
\begin{eqnarray}
\bar n = 32 \int_0^{\infty} d \bar l \, \left ( 2 \bar l^2 - \bar \theta^2 \right)
\left [- \frac{\partial f_0(\bar E)}{\partial \bar E} \right ].
\end{eqnarray}
Using the transformation described in appendix ~\ref{appa}, the density can be expressed as an integration in the energy.
\begin{eqnarray} \label{numpar0}
\bar n = \int_{\bar \Delta}^{\infty}d\bar E \,
\bar E \sqrt{\bar E^2-\bar \Delta^2}\left [- \frac{\partial f_0(\bar E)}{\partial \bar E} \right ] \nonumber \\
\end{eqnarray}
The zero temperature is straightforwardly obtained since  only states in the Fermi surface contribute.
\begin{eqnarray}
-\frac{\partial f_0(\bar E)}{\partial \bar E}\vert_{T=0} = \delta(\bar E- \bar E_F), \;\bar E_F=\bar \mu,
\end{eqnarray}
which renders that for $T=0$ the density of particles is,
\begin{eqnarray}
\bar n =  \,\bar \mu \sqrt{\bar \mu^2-\bar \Delta^2}.
\end{eqnarray}

\subsection{The electrical conductivity}\label{elecsubsec}
To obtain the electrical conductivity we apply  the standard Boltzmann-BGK framework~\cite{ashcroft76}.
The current density and velocity are given by,
$\vec J = (q/V)\sum_{\vec k}\vec v (\vec k)f(\vec k)$, and $\vec v(\vec k)=(1/\hbar)\partial E(\vec k)/\partial \vec k$.
The equilibrium distribution function is slightly changed by the presence of the applied electric field $\vec E$:
$f(\vec k) \approx f_0(\vec k) -(q/\hbar)\tau \vec E \cdot \partial f_0(\vec k)/\partial \vec k$. Since $\sum_{\vec k}\vec v (\vec k)f_0(\vec k)=0$ the current density becomes,
\begin{eqnarray}
\vec J = - \frac{q^2}{\hbar}\tau \frac{1}{V}\sum_{\vec k} \vec E \cdot \frac{\partial f_0(\vec k)}{\partial \vec k}.
\end{eqnarray}
The conductivity is diagonal, $\sigma_{i\;j}=\delta_{i\;j}\sigma$, and equal to,
\begin{eqnarray}\label{electrcond}
\sigma = \frac{q^2}{2V}\tau \sum_{\vec k} \vec v(\vec k)^2 \left [- \frac{\partial f_0(E(\vec k))}{\partial E(\vec k)} \right ],
\end{eqnarray}
as $\vec J = \sigma \vec E$ instead of $J_i = \sigma_{i\;j} E_j$, $i,j=1,2$.

This is because the dispersion relation of Eq.(\ref{disprel}) gives for the velocity,
\begin{eqnarray}
\vec v(\vec k) = \frac{\hbar}{m L_3} \frac{2q^2-\theta^2}{q^3}\vec k,
\end{eqnarray}
since $\partial f_0(\vec k)/\partial \vec k=\hbar \vec v(\vec k) \, \partial f_0(E)/\partial E$ and inside the sum it holds that $v_i(\vec k)v_j(\vec k)=\delta_{i\,j} \vec v(\vec k)^2/2$.
One obtains the dimensionless electrical conductivity from Eq.(\ref{electrcond}),
\begin{eqnarray}
\bar \sigma = \frac{1}{2\pi}\int_0^{\infty}d\bar l \, \bar l \left( \bar l^2+ \bar \theta^2\right)
\left( \frac{2\bar l^2- \bar \theta^2}{\bar l^3}\right)^2\left [- \frac{\partial f_0(\bar E)}{\partial \bar E} \right ]\nonumber \\
\end{eqnarray}
using that $\sum_{\vec k}= (L/2\pi)^2\int d^2 \vec k =(L^2/2\pi)\int_{\theta}^{\infty} dk\,k=(L^2/2\pi L_3^2)\int_{0}^{\infty} d\bar l\, \bar l$. Next we express it as an integration in energy, using the results of Appendix \ref{appa}.
\begin{eqnarray}
&&\bar \sigma = \frac{1}{8\pi}\int_{\sqrt{2}4\bar \theta}^{\infty}d\bar E \,
\sqrt{\bar E^2-2(4\bar \theta)^2} \nonumber \\
&&\left ( \frac{\bar E \bar l_{+}+2\bar \theta^2}{\bar E \bar l_{+}-2\bar \theta^2}-\frac{\bar E \bar l_{-}+2\bar \theta^2}{\bar E \bar l_{-}-2\bar \theta^2}\right)\left [- \frac{\partial f_0(\bar E)}{\partial \bar E} \right ]
\end{eqnarray}
This expression can be expressed without any loss or addition of content as,
\begin{eqnarray}\label{barsig}
&&\bar \sigma = \frac{1}{\pi\bar \Delta^2}\int_{\bar \Delta}^{\infty}d\bar E \,
\sqrt{\bar E^2-\bar \Delta^2} \nonumber \\
&& \left (\bar E^2-\frac{1}{4}\bar \Delta^2 \right ) \left [- \frac{\partial f_0(\bar E)}{\partial \bar E} \right ].
\end{eqnarray}
Remarkably a factor $1/\bar \Delta^2$ pops out of the integral and this holds even in case the collision time depends on the energy, $\tau(E)$. This is a direct consequence of the two (inner and outer) branches of the Fermi surface  that contribute to the conductivity.
Hence from the above expression the collision time can be redefined as $\tau/\bar \Delta^2$ at any temperature.
Consequently in the ZHS limit, which corresponds to $\bar \Delta \rightarrow 0$ ($\theta \rightarrow 0$), the regime  is necessarily ballistic. \\

The  $T=0$ electrical conductivity is straightforwardly obtained.
\begin{eqnarray}
\bar \sigma = \frac{1}{\pi\bar \Delta^2} \sqrt{\bar \mu^2 - \bar \Delta^2} \left (\bar \mu^2 - \frac{1}{4}\bar \Delta^2 \right)
\end{eqnarray}
It can be expressed in terms of the density $\bar n$ instead of the chemical potential $\bar \mu$.
\begin{eqnarray}\label{barsigt0}
&& \bar \sigma(\bar n, \bar \Delta) = \frac{1}{2 \pi \sqrt{2}\bar \Delta^2 } \left [\sqrt{\bar \Delta^2+4 \bar n^2} - \bar \Delta^2 \right ]^{1/2}  \nonumber \\
&&\left [\sqrt{\bar \Delta^2+4 \bar n^2} +\frac{1}{4} \bar \Delta^2 \right ]
\end{eqnarray}
\begin{figure}[ht]
\center
\includegraphics[width=\columnwidth]{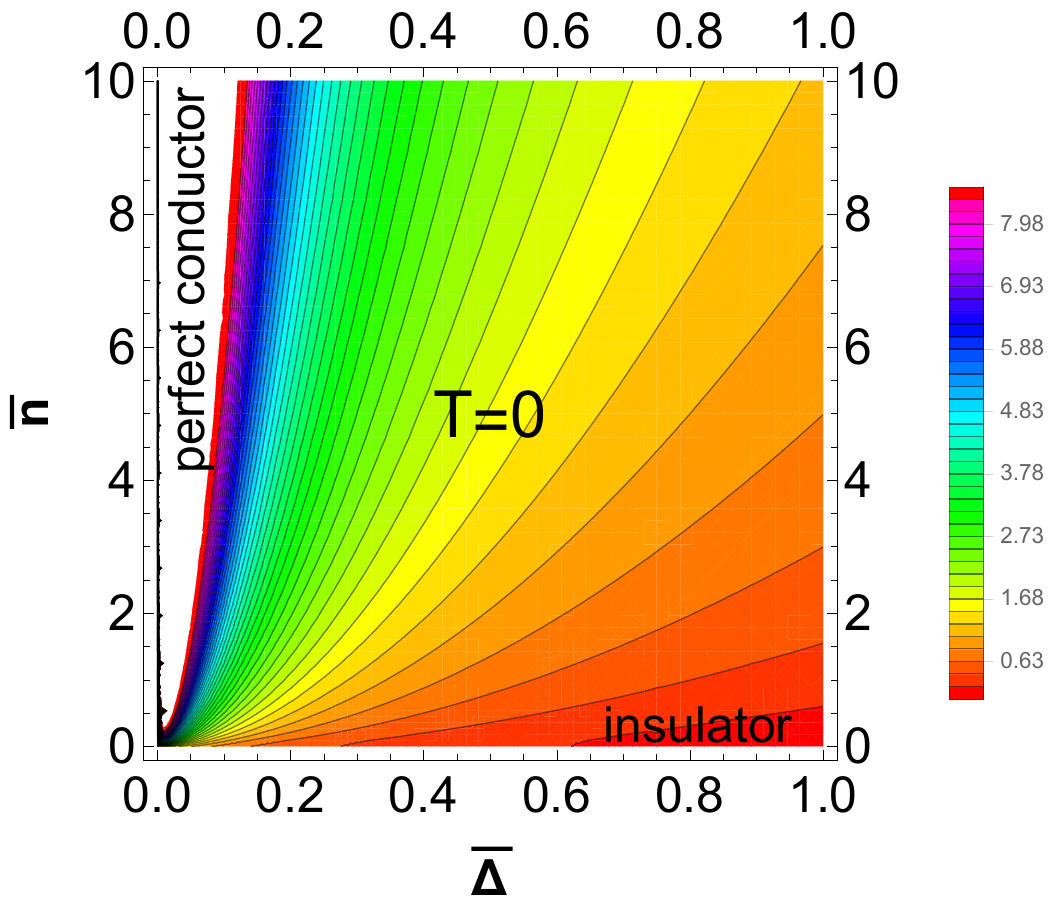}
\caption{The conductivity $\bar \sigma(\bar n, \bar \Delta)$ at $T=0$ versus the gap, $\bar \Delta$ and the density $\bar n$, as given by Eq.(\ref{barsigt0}). The contour lines of $\bar \sigma$ are shown ranging from an insulator ($\bar \sigma=0$, $\bar n=0$) to a perfect conductor ($\bar \sigma=\infty$, $\bar \Delta=0$).}
\label{figdensn}
\end{figure}
Fig.~\ref{figdensn} shows that the DSTM predicts a conductivity ranging from an insulator to a perfect conductor as described by the above $T=0$ formula. Notice the presence of a conductor-insulator transition since the electrical conductivity vanishes ($\bar \sigma \rightarrow 0$) by tuning the chemical potential to the bottom of the conducting band, namely, for
$\bar \mu \rightarrow \bar \Delta$. For a given gap, $\bar \Delta$, the state goes from an insulator, $\bar \sigma \rightarrow 0 $ for $\bar n \rightarrow 0$, to a conductor, $\bar \sigma \rightarrow \bar n^{3/2}/\pi \bar \Delta^2 $ for $\bar n \gg  \bar \Delta$, by adjustment of the density $\bar n$. For $\bar \Delta \rightarrow 0$, namely, in the Dirac linear spectrum, a perfect conductor regime is reached since $\bar \sigma \rightarrow \infty $.

\subsection{The thermal conductivity}\label{thersubsec}
The thermal conductivity based on the Boltzmann-BGK equation follows for the thermal gradient $\vec \nabla T$ correction to the Fermi-Dirac distribution function, namely,
$f(\vec k) \approx f_0(\vec k) -\tau \vec v\cdot  \vec \nabla T\partial f_0(\vec k)/\partial T$.
The first principle of thermodynamics in the absence of external work states that heat $Q$ is affected by the internal energy $E$ and the number of particles $N$, namely, $\delta Q = dE - \mu dN$. In terms of the current density this means that $\vec J_Q = \vec J_E - \mu \vec J_N$ and this results in $\vec J_Q = (1/V) \sum_{\vec k} [E(\vec k)-\mu  ]\vec v(\vec k) f(\vec k)$.
Similar considerations used in case of the electrical conductivity also apply here. Only the correction to the equilibrium distribution $f_0(\vec k)$ yields a current and there are no off diagonal contributions to $\vec J_Q$ such that the expression $v_i(\vec k)v_j(\vec k)=\delta_{i\,j} \vec v(\vec k)^2/2$ is used inside the sum. Thus one obtains that $\vec J_Q = -\kappa \vec \nabla T$. Since $\partial f_0/\partial T = - [(E-\mu)/T] \partial f_0/ \partial E$, one obtains for the thermal conductivity:
\begin{eqnarray}\label{thermcond}
\kappa = \frac{1}{2V}\frac{\tau}{T} \sum_{\vec k} \left [ E(\vec k)-\mu \right ]^2\vec v(\vec k)^2 \left [- \frac{\partial f_0(E(\vec k))}{\partial E(\vec k)} \right ].
\end{eqnarray}
The dimensionless thermal conductivity acquires an expression similar to Eq.(\ref{barsig}).
\begin{eqnarray}\label{barkap}
&& \bar \kappa =  \frac{\bar \beta^2}{\pi \bar \Delta^2}\int_{\bar \Delta}^{\infty}d\bar E \,
\sqrt{\bar E^2-\Delta^2}\left (
\bar E^2-\frac{1}{4}\bar \Delta^2 \right) \nonumber \\
&& (\bar E - \bar \mu)^2 \left [- \frac{\partial f_0(\bar E)}{\partial \bar E} \right ]
\end{eqnarray}
Notice the factor $1/\bar \Delta^2$ that holds even in case the collision time depends on the energy, $\tau(E)$ and renormalizes the collision time.
\subsection{The conductivities and the number of particles in the linear Dirac spectrum }\label{lindirsubsec}
The dimensionless number of particles and conductivities are given by,
\begin{eqnarray}
&&\bar n = \int_{-\bar \beta (\bar \mu-\bar \Delta)}^{\infty}dz \,\bar E(z)
\sqrt{\bar E(z)^2-\bar \Delta^2}\frac{e^z}{\left ( e^z+1\right )^2},
\end{eqnarray}
\begin{eqnarray}
&&\bar \sigma = \frac{1}{\pi \bar \Delta^2}\int_{-\bar \beta (\bar \mu-\bar \Delta)}^{\infty}dz \,
\sqrt{\bar E(z)^2-\bar \Delta^2} \nonumber \\
&&\left ( \bar E(z)^2-\frac{1}{4}\bar \Delta^2\right)\frac{e^z}{\left ( e^z+1\right )^2},\nonumber \\
\end{eqnarray}
and
\begin{eqnarray}
&&\bar \kappa = \frac{1}{\pi \bar \Delta^2}\int_{-\bar \beta (\bar \mu-\bar \Delta)}^{\infty}dz \,
\sqrt{\bar E(z)^2-\bar \Delta^2} \nonumber \\
&&\left ( \bar E(z)^2-\frac{1}{4}\bar \Delta^2\right)z^2\frac{e^z}{\left ( e^z+1\right )^2},\nonumber \\
\end{eqnarray}
where $\bar E(z) = z/\bar \beta +\bar \mu $.
The variable $z \equiv \bar \beta (\bar E- \bar \mu)$ is introduced to treat the ZHS limit ($\bar \Delta\rightarrow 0$).
\begin{eqnarray}
 f_0(\bar E) = \frac{1}{ e^z+1}, \quad  - \frac{\partial f_0(\bar E)}{\partial \bar E} \, d\bar E = \frac{e^z}{\left ( e^z+1\right )^2} \, dz
\end{eqnarray}
Then it becomes easy to take the leading order in the above expressions for $\bar \Delta\rightarrow 0$.
\begin{eqnarray} \label{nbetamu}
\bar n = \frac{1}{\bar \beta^2} \int_{-\bar \beta \bar \mu}^{\infty}dz \,
\left ( z+ \bar \beta \bar \mu \right )^2 \frac{e^z}{\left ( e^z+1\right )^2},
\end{eqnarray}
\begin{eqnarray}
\bar \sigma = \frac{1}{\pi \bar \Delta^2\, \bar \beta^3}\int_{-\bar \beta \bar \mu}^{\infty}dz \,
\left ( z+ \bar \beta \bar \mu \right )^3 \frac{e^z}{\left ( e^z+1\right )^2},
\end{eqnarray}
and
\begin{eqnarray}
\bar \kappa = \frac{1}{\pi \bar \Delta^2\, \bar \beta^3}\int_{-\bar \beta \bar \mu}^{\infty}dz \,
\left ( z+ \bar \beta \bar \mu \right )^3 z^2\frac{e^z}{\left ( e^z+1\right )^2}.
\end{eqnarray}
\begin{figure}[ht]
\center
\includegraphics[width=\columnwidth]{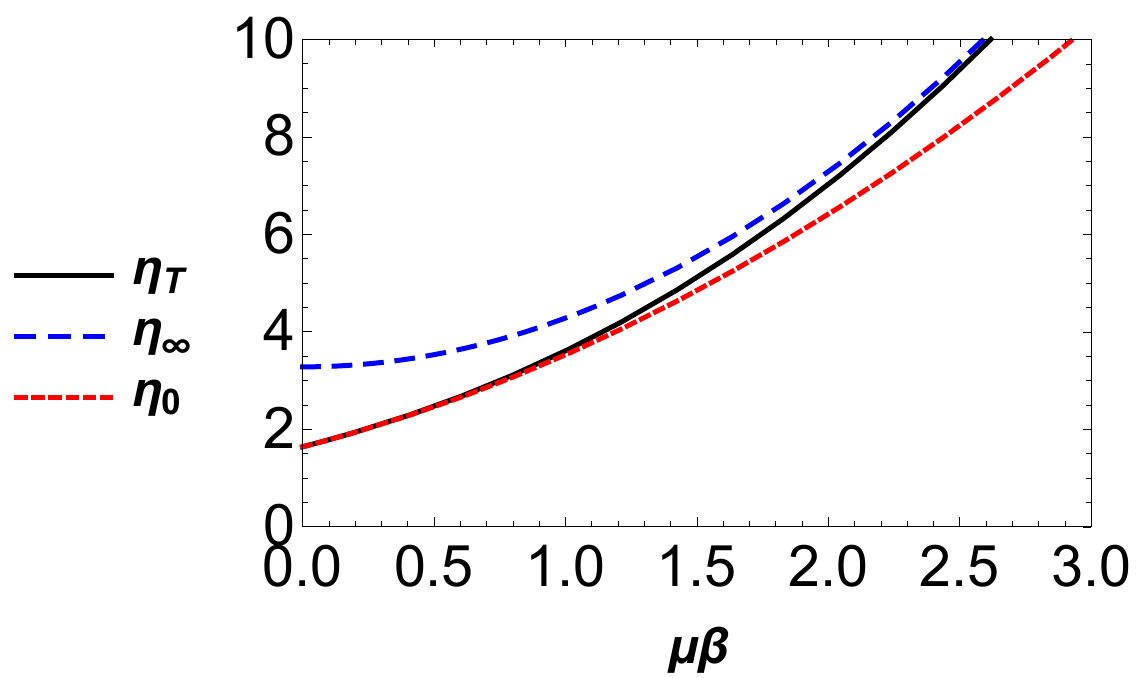}
\caption{The function $\eta_T$, defined by Eq.(\ref{etat}), is displayed here versus the product of the chemical potential, $\mu$, and the inverse temperature, $\beta$.
The functions $\eta_0$ and $\eta_{\infty}$, defined by Eqs.(\ref{etazero}) and (\ref{etainfty}), respectively, are good approximations for
$\eta_T$, below and above the crossover that takes place in the vicinity of $\mu \beta \sim 1.5$.}
\label{figeta}
\end{figure}

\subsection{Conductivities at the low and high temperature limits}\label{lowhigsubsec}
To determine the chemical potential as a function of the temperature and of the Fermi temperature, $\mu(T,T_F)$, the first step is to rewrite Eq.(\ref{nbetamu}) as,
\begin{eqnarray}
&& \left (\frac{T_F}{T} \right)^2 = \eta_T(\beta \mu), \label{etat}\\
&& \eta_T(\beta \mu) \equiv \int_{- \beta \mu}^{\infty}dz \,
\left ( z+  \beta \mu \right )^2 \frac{e^z}{\left ( e^z+1\right )^2}. \nonumber \\
\end{eqnarray}
Analytical expressions follow by approximating the integration limit either as
$\int_{- \beta \mu}^{\infty} \rightarrow \int_{- \infty}^{\infty}$ or as
$\int_{- \beta \mu}^{\infty} \rightarrow \int_{0}^{\infty}$. Hereafter we call these approximations as low ($T \rightarrow 0$, $\beta \rightarrow \infty$ ) and high ($T \rightarrow \infty$, $\beta \rightarrow 0$ ) temperature limits, respectively. Their meaning is discussed below.
We find useful to define the two auxiliary functions,
\begin{eqnarray}
&& \eta_{0}(\beta \mu) \equiv \int_{- \infty}^{\infty}dz \,
\left ( z+  \beta \mu \right )^2 \frac{e^z}{\left ( e^z+1\right )^2}= \nonumber \\
&& I_2+I_0\, \left( \beta \mu \right )^2  \label{etazero} \\
&& \eta_{\infty}(\beta \mu) \equiv \int_{0}^{\infty}dz \,
\left ( z+  \beta \mu \right )^2 \frac{e^z}{\left ( e^z+1\right )^2} =\nonumber \\
&& J_2+2 J_1\left( \beta \mu \right ) + J_0\, \left( \beta \mu \right )^2, \label{etainfty}
\end{eqnarray}
where the integrals $I_n$ and $J_n$ have been find in the appendix ~\ref{appb}. In Fig.~\ref{figeta} the three functions, $\eta_T(\beta \mu)$, $\eta_{0}(\beta \mu)$, and $\eta_{\infty}(\beta \mu)$ are plotted versus $\beta \mu$ to show the range of the validity of such approximations with respect to the true function.
We clearly see a crossover between the regimes. The function $\eta_{0}(\beta \mu)$ is a fair approximation for large $\beta \mu$ up to a lower value and, similarly, $\eta_{\infty}(\beta \mu)$ works in the small $\beta \mu$ region up to an upper value. This crossover value corresponds to $\beta \mu \sim 1.5$.\\

The low temperature limit is set by the equation $(T/T_F)^2= \eta_0(\beta \mu)$, whose solution is,
\begin{eqnarray}
\mu(T) = E_F \sqrt{ 1 - \frac{\pi^2}{6}\left( \frac{T}{T_F}\right)^2}
\end{eqnarray}
where we have used the values for the integrals given in appendix~\ref{appb}.
The high temperature limit is set by the equation $(T/T_F)^2= \eta_{\infty}(\beta \mu)$, whose solution is,
\begin{eqnarray}
&&\mu(T) = 2 \ln{2} E_F \left ( \frac{T}{T_F}\right) \nonumber \\
&&\left \{\sqrt{ 1 + \frac{1}{2 \ln^2{2}}\left [ \left(\frac{T_F}{T}\right)^2- \frac{\pi^2}{6}\right]}-1 \right \}.
\end{eqnarray}

The chemical potential approaches  zero for $T \sim T_{max}$, and is given by,
\begin{eqnarray}
\mu(T) = \frac{1}{2 \ln{2}} E_F \left ( \frac{T}{T_F}\right) \left [ \left(\frac{T_F}{T}\right)^2- \frac{\pi^2}{6}\right].
\end{eqnarray}
This maximum temperature is slightly below the Fermi temperature, and for $ T > T_{max}$  the linear  spectrum regime breaks down.
\begin{eqnarray}
T_{max}= \frac{\sqrt{6}}{\pi} T_F \approx 0.78 T_F.
\end{eqnarray}.
\\

The electrical and thermal conductivity have analytical expressions in the above defined low ($ T \ll T_{F}$) and high ($T \sim T_{max}$) temperature regimes. For the low temperature regime  the lower integration limit is fixed to $-\infty$:
\begin{eqnarray}
&& \bar \sigma = \frac{1}{\pi \bar \Delta^2\, \bar \beta^3}\int_{-\infty }^{\infty}dz \,
\left ( z+ \bar \beta \bar \mu \right )^3 \frac{e^z}{\left ( e^z+1\right )^2} \nonumber \\
&& = \frac{1}{\pi \bar \Delta^2\, \bar \beta^3}\left [3 I_2 (\bar \beta \bar \mu) +I_0 (\bar \beta \bar \mu)^3\right ]\\
&& \bar \kappa = \frac{1}{\pi \bar \Delta^2\, \bar \beta^3}\int_{-\infty }^{\infty}dz \,
\left ( z+ \bar \beta \bar \mu \right )^3 z^2 \frac{e^z}{\left ( e^z+1\right )^2} \nonumber \\
&&  = \frac{1}{\pi \bar \Delta^2\, \bar \beta^3}\left [3 I_4 (\bar \beta \bar \mu) +I_2 (\bar \beta \bar \mu)^3\right ].
\end{eqnarray}
Hence the lowest correction in temperature for the conductivities are given by,
\begin{eqnarray}
&& \bar \sigma = \frac{1}{\pi}\left (\frac{\bar E_F}{\bar \Delta} \right )^3
\left [1+\frac{\pi^2}{2} \left( \frac{T}{T_F}\right)^2\right ]  \\
&& \bar \kappa = \frac{1}{\pi}\left (\frac{\bar E_F}{\bar \Delta} \right )^3 \, \frac{\pi^2}{3}
\left [1+3.7\pi^2 \left( \frac{T}{T_F}\right)^2\right ].
\end{eqnarray}

The high temperature regime is obtained by setting the lower integration limit to $0$:
\begin{eqnarray}
&& \bar \sigma = \frac{1}{\pi \bar \Delta^2\, \bar \beta^3}\int_{0}^{\infty}dz \,
\left ( z+ \bar \beta \bar \mu \right )^3 \frac{e^z}{\left ( e^z+1\right )^2} \\
&& = \frac{1}{\pi \bar \Delta^2\, \bar \beta^3}\left [J_3+ 3 J_2 (\bar \beta \bar \mu) +3 J_1 (\bar \beta \bar \mu)^2 +J_0 (\bar \beta \bar \mu)^3\right ],\nonumber \\
&& \bar \kappa = \frac{1}{\pi \bar \Delta^2\, \bar \beta^3}\int_{0}^{\infty}dz \,
\left ( z+ \bar \beta \bar \mu \right )^3 z^2 \frac{e^z}{\left ( e^z+1\right )^2} \\
&& = \frac{1}{\pi \bar \Delta^2\, \bar \beta^3}\left [J_5+ 3 J_4 (\bar \beta \bar \mu) +3 J_3 (\bar \beta \bar \mu)^2 +J_2 (\bar \beta \bar \mu)^3\right ].\nonumber \\
\end{eqnarray}
In this temperature range the conductivities approach the asymptotic limit,
\begin{eqnarray}
&& \bar \sigma \rightarrow \frac{1}{\pi}\left (\frac{\bar E_F}{\bar \Delta} \right )^3
\left( \frac{T}{T_F}\right)^3J_3, \\
&& \bar \kappa \rightarrow \frac{1}{\pi}\left (\frac{\bar E_F}{\bar \Delta} \right )^3
\left( \frac{T}{T_F}\right)^3J_5.
\end{eqnarray}

\subsection{The Wiedemann-Franz law at the low and high temperature limits}\label{wiedsubsec}
The above obtained conductivities increase with respect to the temperature as they do not take into account any temperature or phononic contributions to the collision time $\tau$. For this reason it is interesting to obtain the ratio between the dimensionless thermal and electrical conductivity since it determines the deviation from the standard Wiedemann-Franz law, according to  Eq.(\ref{kapsig0}).
One obtains that in the low temperature regime, $T \ll T_F$, the  Wiedemann-Franz law is satisfied,
\begin{eqnarray}
\frac{\bar \kappa}{\bar \sigma} = \frac{\pi^2}{3}
\left [1+3.2\pi^2 \left( \frac{T}{T_F}\right)^2\right ],
\end{eqnarray}
while in the limit $T \sim T_{max}$ the ratio saturates in the limit,
\begin{eqnarray}
\frac{\bar \kappa}{\bar\sigma} \rightarrow \frac{J_5}{J_3}=6.5552 \;\frac{\pi^2}{3},
\end{eqnarray}
The temperatures $T_F$ and $T_{max}$ are very sensitive to the density and the velocity parameters and can fall into experimentally reachable bounds.
For an electronic density and velocity of $n = 10^{12}\;\mbox{cm}^{-2}$ and $v_0 = 10^5\; \mbox{m/s}$, respectively $T_F = 2.64 \; 10^{3}\; \mbox{K}$. They correspond to  $k_F=3.5 \; 10^{-1} \; \mbox{nm$^{-1}$}$ and $L_3 = 2.0 \; \mbox{nm}$. At room temperature, $T = 300 \; \mbox{K}$, $T/T_F = 0.11$, which gives $\bar \kappa/\bar \sigma = 1.38 \; \pi^2/3$. This  is a noticeable deviation from the standard Wiedemann-Franz law~\cite{fong13}. Next consider the low density $n = 10^{9}\;\mbox{cm}^{-2}$ ($k_F=1.1 \; 10^{-2} \; \mbox{nm}$), and  $v_0$ as before.
Then the Fermi temperature is much lower, $T_F = 83 \; \mbox{K}$, and so, $T_{max}= 65 \; \mbox{K}$. Therefore for such a low density the above predicted saturation limit is experimentally reachable.

\section{Inclusion of magnetic field effects}\label{magsec}
Remarkably the magnetic energy is attractive among the particles in the linear Dirac spectrum limit~\cite{doria17}.
Although this attractiveness will not be considered here, as the magnetic energy is proven to be very small, it is interesting to show its existence~\cite{doria17}.
To see it cast the original hamiltonian of Eq.(\ref{dseq}) with the three term kinetic energy included.
\begin{eqnarray}\label{dseq2}
&& H = \int d^3x \; \left \{ \frac{1}{2m}\vert\vec{\sigma}\cdot\vec P\Psi \vert^2-\frac{\hbar}{4m}\vec \nabla \cdot \big [\Psi^\dag\left(\vec \sigma \times \vec P\right)\Psi  \right .\nonumber \\
&& +c.c. \big ] \left . + \frac{\hbar q }{2 m c}\vec{h} \cdot\left(\Psi^\dag\vec\sigma\Psi\right) +\frac{1}{8\pi} \vec h(\Psi)^2  \right \}.
\end{eqnarray}
The  current associated to the three term decomposition is given below.
\begin{eqnarray}
&& \vec J = \frac{q}{2m}\left [ \left(\vec \sigma \cdot \vec P \Psi\right)^{\dag}\vec \sigma \Psi +c.c \right]-\frac{\hbar q}{2m}\vec \nabla \times  \left( \Psi^{\dag}\vec \sigma \Psi\right).\nonumber \\
\end{eqnarray}
Then it becomes clear that for a ZHS, defined by Eq.(\ref{foe1}), Amp\`ere's law can be exactly solved, and the magnetic field determined, as given by Eq.(\ref{foe2}). Then the hamiltonian of Eq.(\ref{dseq2}) can be expressed as,
\begin{eqnarray}\label{dseq3}
H= v_0\hbar \, \sum_{\vec k,\; k \ne 0} k  \,{c_{\vec k}}^\dag c_{\vec k}-\frac{1}{8\pi}\int d^3x \;\vec{h}^2,
\end{eqnarray}
which shows that the magnetic interaction among the Dirac particles is attractive. As shown before in Ref.~\onlinecite{doria17}, this magnetic energy is residual as it is proportional to $r_e/L$, where $r_e=q^2/mc\approx 2.8 \,  \, 10^{-6}$ nm is the electron's classical radius. \\

From the above results it becomes clear that the accessible energy levels to enter the DSTM are of nearly non interacting linear Dirac states. The single particle states are given by,
\begin{eqnarray}\label{norm0}
|\chi \rangle = \sum_{\vec k}\, f(\vec k) \,c_{\vec k}^{\dag}|0\rangle, \quad \sum_{\vec k}\,|f(\vec k)|^2 \, =1.
\end{eqnarray}
The energy expectation value is straightforwardly obtained.
\begin{eqnarray}
&& \langle \chi| H_0 |\chi \rangle =\sum_{\vec k, k \ne 0}\, k \, |f(\vec k)|^2, \\
&& H_0= v_0\hbar \sum_{\vec k,\, k \ne 0} k  \,{c_{\vec k}}^\dag c_{\vec k}.
\end{eqnarray}
Although the energy has been removed of its magnetic content, the magnetic field that dresses the particle can still be obtained.
\begin{eqnarray}
\langle \vec h \rangle = \langle \chi| \vec h |\chi \rangle = -4\pi \mu_B \, \langle \chi| \Psi^{\dag}\vec \sigma \Psi|\chi \rangle.
\end{eqnarray}
The components are given by,
\begin{eqnarray}
&& \langle h_1 \rangle =-4\pi \mu_B\, i\frac{x_3}{\vert x_3\vert} \left (  b^* a- b a^*\right ),\\
&& \langle h_2 \rangle =4\pi \mu_B\, \frac{x_3}{\vert x_3\vert} \left (  b^* a_0+ b a^*\right ),\\
&& \langle h_3 \rangle =-4\pi \mu_B\, \left ( \vert a\vert^2 - \vert b\vert^2 \right ),
\end{eqnarray}
where we define the following auxiliary functions.
\begin{eqnarray}
&& a = \frac{1}{\sqrt{V}} \sum_{\vec k,\; k >0} f(\vec k) \; e^{i\vec{k}\cdot\vec{x}} e^{-k \vert x_3 \vert  }, \\
&& b = \frac{1}{\sqrt{V}} \sum_{\vec k,\; k >0} \frac{k_{+}}{k} f(\vec k) \; e^{i\vec{k}\cdot\vec{x}} e^{-k \vert x_3 \vert  }.
\end{eqnarray}
At this point it is also possible to verify the previously given argument showing that the magnetic field form closed loops that pierce the layer twice. The in-plane component, $\langle h_1 \hat x_1+ h_2 \hat x_2 \rangle $,  flips sign from above to below the layer, but not the $h_3$ component, and it holds that $\langle \vec \nabla \cdot \vec h \rangle=0$.\\

To study the topological properties a collective state made of single particle states is selected, such that its energy is $v_0\hbar k_0$, and it has zero momentum (\, $\langle \chi| \vec P |\chi \rangle=0$, $\vec P= \hbar \sum_{\vec k,\, k \ne 0}\, \vec k  \,{c_{\vec k}}^\dag c_{\vec k}$\,). An overall momentum can be added to this collective state a posteriori to set it in motion. Thus the convenient choice is to take that $f(\vec k)\equiv f(\vert \vec k \vert)$ and for simplicity the phase is removed, $f^*(\vec k)=  f( \vec k )$. This simplifies substantially the expressions for the local magnetic field that dresses this state.
\begin{eqnarray}
&& \langle h_1 \rangle =-8\pi \mu_B\,\frac{x_3}{\vert x_3\vert} \frac{x_1}{\vert \vec x \vert} a_0 a_1,\\
&& \langle h_2 \rangle =-8\pi \mu_B\, \frac{x_3}{\vert x_3\vert} \frac{x_2}{\vert \vec x \vert} a_0 a_1,\\
&& \langle h_3 \rangle =-4\pi \mu_B\, \left ( a_0^2 - a_1^2 \right ),
\end{eqnarray}
where $\vec x \equiv x_1 \hat x_1 + x_2 \hat x_2$,  $ \vert \vec x \vert \equiv \sqrt{\vec x \cdot \vec x}$ and
\begin{eqnarray}
&& a_0 = \frac{L}{2\pi \sqrt{L_3}} \int_0^{\infty} dk\,k\,e^{-k \vert x_3 \vert  }\,J_0(k\vert \vec x \vert ) f(k), \\
&& a_1 = \frac{L}{2\pi \sqrt{L_3}} \int_0^{\infty} dk\,k\,e^{-k \vert x_3 \vert  }\,J_1(k\vert \vec x \vert ) f(k).
\end{eqnarray}
In order to explicitly  compute the topological number, the state is assumed to have a very small width, defined by $\eta$, around its energy $v_0\hbar k_0$.
Under this simplification an explicit expression for $f(k)$ is proposed to automatically  satisfy the normalization condition, Eq.(\ref{norm0}), in the continuum.
\begin{eqnarray}
&& \frac{L^2}{4\pi}\int_{0}^{\infty}dk^2 \, \vert f(\vec k)\vert^2=1 \rightarrow \vert f(\vec k)\vert^2 = \frac{4\pi}{L^2} \delta_{\eta}\left ( k^2-k_0^2\right),\nonumber \\
&& \delta_{\eta}\left ( k^2-k_0^2\right)= \lim_{\eta \approx 0}
\frac{1}{\sqrt{2\pi\eta}}e^{-\left (k^2-k_0^2 \right)^2/2\eta}. \label{defeta}
\end{eqnarray}
Notice that $\eta$ has the dimension of $k^4$. Under the approximation that Dirac's delta function is given by the gaussian representation, the integrals can be solved approximately. Thus by taking that,
\begin{eqnarray}
 f(k) = \frac{2\sqrt{2\pi}}{L} \left ( 2\pi \eta \right)^{1/4} \delta_{2\eta}\left ( k^2-k_0^2\right),
\end{eqnarray}
the above defined functions are simplified further on.
\begin{eqnarray}
&& a_0 = \left(\frac{\eta}{2\pi} \right)^{1/4}\frac{1}{\sqrt{L_3}} \, e^{-k_0 \vert x_3 \vert  }\,J_0(k_0\vert \vec x \vert ),\\
&& a_1 = \left(\frac{\eta}{2\pi} \right)^{1/4}\frac{1}{\sqrt{L_3}} \, e^{-k_0 \vert x_3 \vert  }\,J_1(k_0\vert \vec x \vert ).
\end{eqnarray}
From the above an expression the magnetic field is obtained and is manageable for the analytical calculation of its Chern-Symons index.
\begin{eqnarray}
&& \langle\vec h \rangle  = -4\pi \mu_B\, \left (\frac{\eta}{2\pi} \right )^{1/2}\frac{1}{L_3}\, e^{-2 k_0 \vert x_3 \vert  }\,\Big \{ 2 \frac{x_3}{\vert x_3\vert}\frac{\vec x}{\vert \vec x \vert} \nonumber  \\
&&J_0(k_0\vert \vec x \vert )J_1(k_0\vert \vec x \vert ) +
\hat x_3 \left [J_0(k_0\vert \vec x \vert )^2-J_1(k_0\vert \vec x \vert )^2 \right ]
\Big \} \nonumber \label{avgh}\\
\end{eqnarray}
At the center ($x_3=0$, $\vert  \vec x \vert =0$) the field is constant and perpendicular to the layer,
\begin{eqnarray}
\langle \vec h_{center} \rangle = -4\pi \mu_B \left (\frac{\eta}{2\pi} \right)^{1/2}\left (\frac{1}{L_3} \right ) \hat x_3.
\end{eqnarray}
Away from the center a magnetic field stream line pierces the layer at several points which means that the field becomes perpendicular to the layer at special radii values. These special radii corresponds to the zeros of the Bessel functions. The field becomes perpendicular  in alternating directions, upward  ($\hat x_3$) for $J_1(y_{n1})=0$, ($y_{n1}=k_0\vert \vec x_{n1} \vert $) and downward ($-\hat x_3$) for $J_0(y_{n0})=0$, ($y_{n0}=k_0\vert \vec x_{n0} \vert$), respectively. \\
\begin{figure}[ht]
\center
\includegraphics[width=\columnwidth]{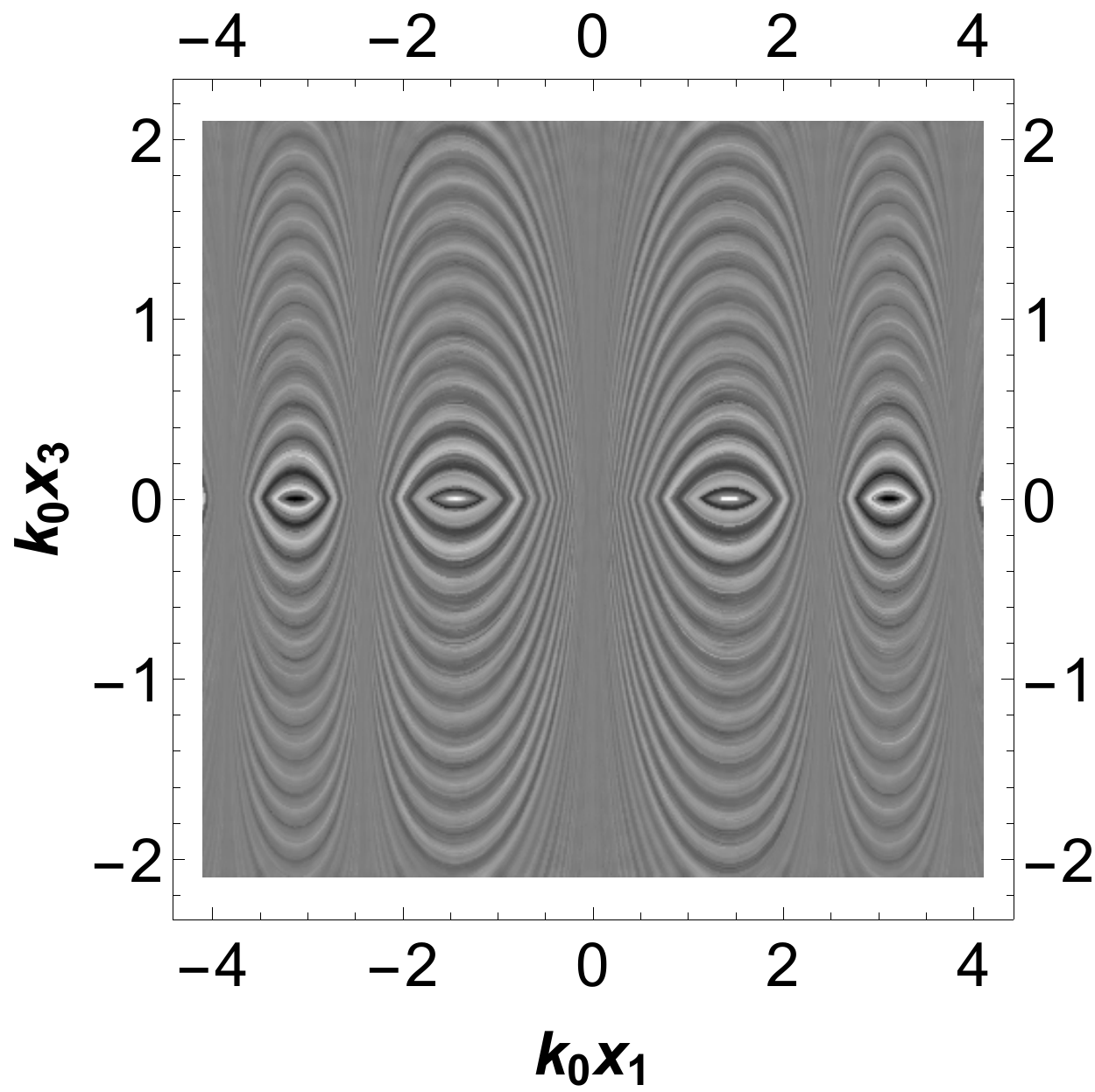}
\caption{A sectional view of the magnetic field that dresses a single particle in the layer, obtained from Eq.(\ref{avgh}), is shown along the plane $(x_1,x_3)$.
Coordinates are scaled by $k_0$ associated to the Dirac linear energy $v_0\hbar k_0$.
The stream lines decay in intensity away from the $x_3=0$ layer according to Eq.(\ref{avgh}).
The schematic view of Fig.~\ref{figstreamline} complements the present one because it provides the direction of the closed loops seen here.
Fig.~\ref{figcont} is also complementary to the present one since it shows the contour lines of this field in the  $x_3=0$ layer.}
\label{figfield}
\end{figure}
%
\begin{figure}[ht]
\center
\includegraphics[width=\columnwidth]{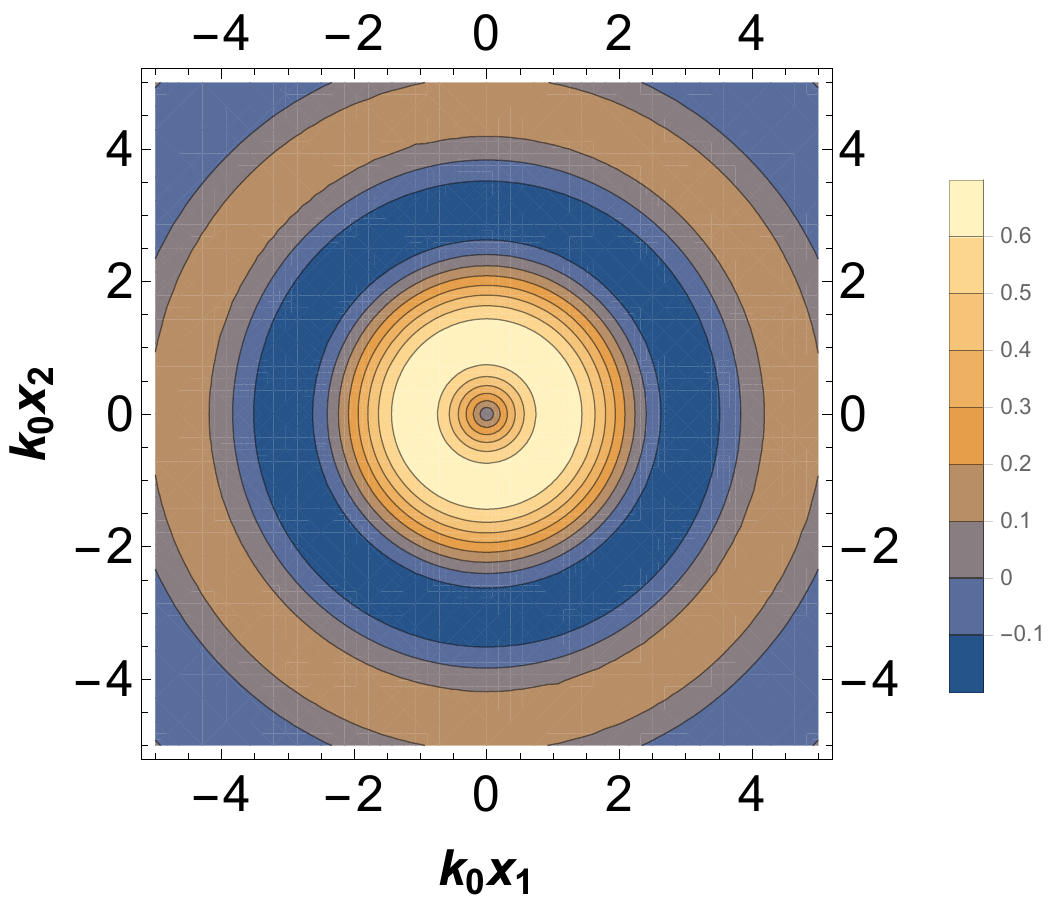}
\caption{The contour lines of the magnetic field that dresses a single particle in the layer, obtained from Eq.(\ref{avgh}), are shown here in the layer ($x_3=0$). The  concentric rings show alternate upward and downward field directions in agreement with the schematic pattern of Fig.~\ref{figstreamline}.
The field is in arbitrary units and coordinates $(x_1,x_2)$ are multiplied by $k_0$ associated to the energy state $v_0\hbar k_0$. The sectional view shown in Fig.~\ref{figfield} is complementary to the present one and allow to envisage the three-dimensional arrangement of the magnetic field.}
\label{figcont}
\end{figure}

Figs.~\ref{figstreamline}, ~\ref{figfield} and ~\ref{figcont}  show this magnetic field whose single particle energy is centered in $v_0\hbar k_0$ with a very small wavenumber spread $\eta^{1/4}$ around the wavenumber $k_0$ as described by Eq.(\ref{defeta}).
Fig.~\ref{figstreamline} shows a pictorial view of the field where the streamlines weave the layer forming packed closed streamlines of alternating circulation. The pictorial view is a helpful tool to understand Figs.~\ref{figfield} and ~\ref{figcont} which show two distinct views of the magnetic field obtained from Eq.(\ref{avgh}). Fig.~\ref{figfield} shows the strength of the field within a sectional view of the plane $(x_1,x_3)$ whereas Fig.~\ref{figcont} shows the contour lines of this field in the layer ($x_3=0$ plane).
\\

There should be no net magnetic flux crossing the layer as there is no external field applied to it. Therefore one must impose that
\begin{eqnarray}
\Phi \equiv \int_{x_3=0^+} d^2\vec x \, \langle  h_3 \rangle =0
\end{eqnarray}
Using the Bessel function identity $\int dx \, x \, [J_0(x)^2-J_1(x)^2]=x J_0(x)J_1(x)$ and integrating over a radius $R=k_0\vert \vec x \vert $ one obtains that the zero flux condition selects that the field at the edge of the unit cell be perpendicularly to the layer.
\begin{eqnarray}
 && J_0( y_{n0})=0  \rightarrow k_0\frac{L}{2}=y_{n0}, \label{regraj0}\\
 && J_1(y_{n1} ) =0 \rightarrow k_0\frac{L}{2}=y_{n1}. \label{regraj1}
\end{eqnarray}
Notice that such condition is the same one of no superficial current $\vec J_s$  circulating at the edge. Amp\`ere´s law implies in the boundary condition $\hat n \times \big(\vec h(0^+ )- \vec h(0^- )\big )=4\pi \vec J_s/c$ and from it one obtains that,
\begin{eqnarray}
\vec J_s= - c \mu_B \left (\frac{\eta}{2\pi} \right)^{1/2}\frac{4}{L_3} J_0(k_0\vert \vec x \vert )J_1(k_0\vert \vec x \vert )\, \hat \phi.
\end{eqnarray}
Thus the conditions for no flux, $\Phi=0$, also guarantee no current circulating at the edge, $\vec J_s=0$. \\

Finally the topological number $Q$  (Chern-Simons index), defined by Eq.(\ref{topo}), is analytically obtained in the appendix \ref{appc} and given by,
\begin{eqnarray}\label{chernvalue0}
Q=-\frac{1}{2}\left [
\frac{J_0(k_0L/2 )^2-J_1(k_0L/2 )^2}{J_0(k_0L/2 )^2+J_1(k_0L/2 )^2}
-1 \right ].
\end{eqnarray}
In this respect the two possible zero flux conditions result in very distinct results,
\begin{eqnarray}
&& Q =-1 \, \mbox{for} \, k_0\frac{L}{2}=y_{n0}, \\
&& Q =\;0 \, \mbox{for} \; k_0\frac{L}{2}=y_{n1}.
\end{eqnarray}
Hence there are trivial ($Q=0$) and  non-trivial ($Q=-1$) topological solutions. For the trivial solution the field at the edge points in the same direction of the center whereas for the non-trivial edge and center point in opposite directions.
Notice that it is possible to have also solutions $Q=+1$ by reverting the direction of the currents and the local field. Therefore out of the original Schroedinger hamiltonian two dressed topologically non trivial states are obtained, $Q= \pm 1$.\\

The conditions that set $k_0 L/2$ equals to either $y_{n0}$ or $y_{n1}$ can be interpreted as defining the accessible quantized states in the unit cell area $L^2$. These states are filled to yield a reconstructed Fermi surface.
For instance, in case of the non-trivial states, namely those with $Q=-1$,  only  fill the quantized wavenumber states $k_{n0} = 2 y_{n0}/L$ until the Fermi level is reached. A bettwer understanding is reached by looking  the limit  $k_0 L /2 \gg 1$. In this case we use the approximated formula for the Bessel functions for large argument $y$, namely, $J_0(y) \rightarrow \sqrt{2/(\pi y)} \cos{(y-\pi/4)}$ such that the zeros are $y_{n0}= (n-1/4)\pi$ which gives $k_{n0}=(2n+3/2)\pi/L$. From the other side the topologically trivial states are associated to $J_1(y) \rightarrow \sqrt{2/(\pi y)} \sin{(y-\pi/4)}$, such that the zeros are $y_{n1}= (n+1/4)\pi$ which gives that the states $k_{n1}=(2n+1/2)\pi/L$ are trivial.\\

\subsection{Estimate of the  magnetic field and of the magnetic energy}\label{estimsubsec}
To estimate the magnetic field  of the above single particle states we consider a wavenumber at the Fermi level of the linear Dirac spectrum, $k_0=k_F=\sqrt{4\pi n}$. The width of the state, defined in Eq.(\ref{defeta}), should be a small fraction of the Fermi wavenumber, $\eta = (\varepsilon k_F)^{4}$, and this smallness is given by  dimensionless number $\varepsilon$. In this case, the field at the center is expressed as $\langle h_{center} \rangle = 4 (2\pi)^{3/2} \mu_B \varepsilon^2 n/L_3$. To estimate it consider that $\mu_B = 9.2 \, \mbox{G.nm$^3$}$, together with the values previously taken of $n = 10^{12}\;\mbox{cm}^{-2}$ and $L_3 = 2.0 \; \mbox{nm}$ ($v_0 = 10^5\; \mbox{m/s}$). Then one obtains that $\langle h_{center} \rangle =2.9\,\varepsilon^2 \,\mbox{G}$. The choice of width $\varepsilon \sim 10^{-1}$ shows that the field is in the sub-Gauss regime, $\langle h_{center}\rangle \sim 2.9\,\mbox{x}\, 10^{-2}\,\mbox{G}$. The  magnetic field energy is estimated under the approximation $F \equiv \int d^3x \,\vec{h}^2/8\pi \approx \int d^3x \, \langle \vec{h}\rangle^2/8\pi$. Using the expectation value of the magnetic field given by Eq.(\ref{avgh}), one obtains that
$F = 4\pi \varepsilon^4 \mu_B^2 (k_F /L_3^2) I$, $I = \int_0^{L k_F/2} dy y [ J_0^2(y) + J_1^2(y)]^2$.
The scale is set by the square of Bohr's magneton, $\mu_B^2 = 5.3 \, 10^{-8}\,\mbox{eV . nm$^3$}$ which gives $F = 5.8 \,\mbox{x}\, 10^{-8}\,\varepsilon^4 I \, \mbox{eV}$. Therefore the magnetic field energy is very small, $F \sim 10^{-12}\, \mbox{eV}$, in comparison with
the Fermi energy  of the linear Dirac spectrum, $\hbar v_0 k_F = 2.3 \,\mbox{x}\, 10^{-2}\, \mbox{eV}$.\\

For completeness the local magnetic field created by the charged particles in the original Drude and Sommerfeld models is also estimated. Assume that the current created by a single particle in straight motion, between two collisions is $I=q/\tau=1.6\; 10^{-2}$ mA, where $\tau \sim 10^{-14}$ s.
The resulting magnetic field is $h = 43.0$ G and follows from Amp\`ere's law,$ h = \mu_0 I/2\pi r$, as felt by another passing electron located at a position , $r = 1/n^{1/3}=0.46$ nm, defined by the electronic density typical of metals, $n \sim 10^{22}$  e/cm$^3$. The net field is expected to be lower than this value as there are other nearby passing electrons, which create  fields in  other  directions. Nevertheless this field has no topological content and therefore can be averaged to zero.\\

\section{Conclusion} \label{conclusion}
The Drude-Sommerfeld scenario of free particles with residual interactions is able to describe Weyl fermions and the linear Dirac spectrum~\cite{doria17} regardless if the underlying lattice symmetry.
The  Weyl state has lower symmetry than the kinetic energy since it breaks its rotational symmetry because it lives in a layer. This is the key ingredient to trigger the onset of a local magnetic field that dresses the particles. Although weak this field is of fundamental importance because it brings topological stability to the states. Thus the particles can live on a higher kinetic energy state as they are protected from decaying into other states.
The Chern-Simons index of the single particle states is calculated and shown to be non-trivial. The electrical and the thermal conductivities of this  Drude-Sommerfeld topological model  have the collision time renormalized by the inverse of the gap square. In the limit that the gap becomes increasingly small (the linear Dirac spectrum limit), a ballistic regime is reached and the Drude-Sommerfeld topological scenario is well justified.

\textbf{Acknowledgments}:
M.M.D acknowledges helpful discussions with  Andrea Perali, Marco Cariglia, Mohammed El Massalam, Marcello Barbosa da Silva Neto, Rodolpho Ribeiro Gomes, Rodrigo Coelho and Yakov Kopelevich.\\

\appendix
\section{Integration in energy} \label{appa}
The integration of an arbitrary function $F(\bar l)$  in terms of the wavenumber $\bar l$,
\begin{eqnarray}
I = \int_0^{\infty}d\bar l \,F(\bar l),
\end{eqnarray}
has to be express in terms of the energy variable, $\bar E =2 \bar \theta^2/\bar l + 4 \bar l$. The difficulty lies in the fact that the function $\bar E(\bar l)$ is single valued whereas $\bar l(\bar E)$ is not. For a given $\bar E$ there are two possible branches given by,
\begin{eqnarray}
\bar l_{\pm}(\bar E) = \frac{\bar E}{8} \pm \frac{1}{8}\sqrt{\bar E^2-2(4\bar \theta)^2}.
\end{eqnarray}
These two branches meet at the bottom of the band, which corresponds to the energy $\bar E = \sqrt{2} 4 \bar \theta$, and to the wavenumber $\bar l_{\pm}= \sqrt{2}\bar \theta/2$.
Hence one must separate the integration over the two branches.
\begin{eqnarray}
I = \int_0^{\sqrt{2}\bar \theta/2}d\bar l_{-}\, F(\bar l_{-})+\int_{\sqrt{2}\bar \theta/2}^{\infty}d\bar l_{+} \, F(\bar l_{+})
\end{eqnarray}
The change to the energy variable gives that,
\begin{eqnarray}
I = \int_0^{\sqrt{2}\bar \theta/2}
d\bar E \,\frac{d\bar l_{-}}{dE} \,F[\bar l_{-}(E)]
+\int_{\sqrt{2}\bar \theta/2}^{\infty}dE \,\frac{d\bar l_{+}}{dE} \,F[\bar l_{+}(E)].\nonumber \\
\end{eqnarray}
Finally one obtains that the integral done in terms of the energy is,
\begin{eqnarray}
&& I = \int_{\sqrt{2} 4 \bar \theta}^{\infty}d\bar E \, G(\bar E),\\
&& G(\bar E) = \frac{1}{8}\left \{ F[\bar l_{+}(E)]-F[\bar l_{-}(E)]\right \} +  \nonumber \\
&& \frac{1}{8} \frac{\bar E}{\sqrt{\bar E^2-2(4\bar \theta)^2}}   \left \{ F[\bar l_{+}(E)]+F[\bar l_{-}(E)]\right \}.
\end{eqnarray}

\section{Useful integrals} \label{appb}
The following integrals are used for the obtainment of the electrical and thermal conductivities.
\begin{eqnarray}
J_n \equiv \int_{0}^{+\infty} dz \, z^n \, \frac{e^z}{\left( e^z+1\right)^2}
\end{eqnarray}
The first six integrals are known and given by,
$J_0=1/2=0.5$, $J_1=\ln{2} \approx 0.69315$, $J_2=\pi^2/6 \approx 1.64493$, $J_3=9\zeta(3)/2\approx 5.40926$, $J_4=7 \pi^4/30 \approx 22.7288 $, and  $J_5\approx 116.654$.\\
The integrals,
\begin{eqnarray}
I_n \equiv \int_{-\infty}^{+\infty} dz \, z^n \, \frac{e^z}{\left( e^z+1\right)^2},
\end{eqnarray}
satisfy $I_n=2J_n$ for $n$ even and $I_n=0$ for $n$ odd.
\section{Calculation of the  Chern-Symons index} \label{appc}
We analytically determine the Chern-Simons index, given by Eq.(\ref{topo}), for the magnetic field of the single particle state, as defined by Eq.(\ref{avgh}).
Although we assume a periodic tiling of the layer with unit cell area $L^2$, the calculation of the Chern-Simons index assumes a disk of radius $R=L/2$.

The direction of the magnetic field is straightforwardly obtained under the above approximations.
\begin{eqnarray}
&& \hat h = 2 A(k_0\vert \vec x \vert)\frac{x_3}{\vert x_3 \vert}\,\frac{\hat x}{\vert \vec x \vert}+ B(k_0\vert \vec x \vert)\hat x_3 \\
&& A(k_0\vert \vec x \vert) \equiv \frac{J_0(k_0\vert \vec x \vert )J_1(k_0\vert \vec x \vert )}{J_0(k_0\vert \vec x \vert )^2+J_1(k_0\vert \vec x \vert )^2}, \label{defA} \\  && B(k_0\vert \vec x \vert) \equiv \frac{J_0(k_0\vert \vec x \vert )^2-J_1(k_0\vert \vec x \vert )^2}{J_0(k_0\vert \vec x \vert )^2+J_1(k_0\vert \vec x \vert )^2} \label{defB}
\end{eqnarray}

Consider the Bessel functions $J_0(y)$ and $J_1(y)$ and define $y \equiv k_0\vert \vec x \vert$.
Define the derivative  $g'\equiv dg/dy$ such that $J_0'=-J_1$ and $J_1'=J_0-J_1/y$.
Then one can prove that,
\begin{eqnarray}
&&\left (\frac{J_0J_1}{J_0^2+J_1^2}\right)'=\frac{J_0^2-J_1^2}{J_0^2+J_1^2}\left(1 -
\frac{1}{y}\frac{J_0J_1}{J_0^2+J_1^2}\right)\\
&&\left (\frac{J_0^2-J_1^2}{J_0^2+J_1^2}\right)'=-4\frac{J_0 J_1}{J_0^2+J_1^2}\left(1 -
\frac{1}{y}\frac{J_0J_1}{J_0^2+J_1^2}\right).
\end{eqnarray}
In terms of the definition of functions $A(y)$ and $B(y)$, given by Eqs.(\ref{defA}) and (\ref{defB}), the above identities become
\begin{eqnarray}
&& A' = B \left ( 1-\frac{1}{y}A \right),\\
&& B' = -4A \left ( 1-\frac{1}{y}A \right),
\end{eqnarray}
respectively.
From $\partial A/\partial x_i=k_0^2 x_i A'/y$ and $\partial B/\partial x_i=k_0^2 x_i B'/y$ it follows the derivatives of the field unit vector.
\begin{eqnarray}
&& \frac{\partial \hat h_1}{\partial x_1}= 2\frac{1}{\vert \vec x \vert^2}\left ( \frac{x_2^2}{\vert \vec x \vert}\, A + x_1^2 \,k A'\right )\\
&& \frac{\partial \hat h_1}{\partial x_2}= \frac{\partial \hat h_2}{\partial x_1}= 2\frac{x_1x_2}{\vert \vec x \vert^2}\left (  - A + k A'\right )\\
&& \frac{\partial \hat h_2}{\partial x_2}= 2\frac{1}{\vert \vec x \vert^2}\left ( \frac{x_1^2}{\vert \vec x \vert}\, A + x_2^2 \,k A'\right )\\
&& \frac{\partial \hat h_3}{\partial x_1}= 2\frac{x_1}{\vert \vec x \vert} \, k B'\\
&& \frac{\partial \hat h_3}{\partial x_2}= 2\frac{x_2}{\vert \vec x \vert} \, k B'
\end{eqnarray}
Then one obtains that,
\begin{eqnarray}
&& \left ( \frac{\partial \hat h_2}{\partial x_1}\frac{\partial \hat h_3}{\partial x_2}- \frac{\partial \hat h_3}{\partial x_1}\frac{\partial \hat h_2}{\partial x_2} \right ) \cdot \hat h_1 =-4 \frac{x_1^2}{\vert \vec x \vert^3}\,k\,A^2\,B' \\
&& \left ( \frac{\partial \hat h_3}{\partial x_1}\frac{\partial \hat h_1}{\partial x_2}- \frac{\partial \hat h_1}{\partial x_1}\frac{\partial \hat h_3}{\partial x_2} \right ) \cdot \hat h_2 =-4 \frac{x_2^2}{\vert \vec x \vert^3}\,k\,A^2\,B' \\
&& \left ( \frac{\partial \hat h_1}{\partial x_1}\frac{\partial \hat h_2}{\partial x_2}- \frac{\partial \hat h_2}{\partial x_1}\frac{\partial \hat h_1}{\partial x_2} \right ) \cdot \hat h_3 = 4 \frac{1}{\vert \vec x \vert^3}\,k\,A'A\,B'
\end{eqnarray}
Such that the sum of the these three terms is given by,
\begin{eqnarray}
\left ( \frac{\partial \hat h}{\partial x_1}\times \frac{\partial \hat h}{\partial x_2}\right ) \cdot \hat h = \frac{4}{y}\, k^2 A \left (B^2+4 A^2 \right)\left (1-\frac{A}{y} \right ).\nonumber \\
\end{eqnarray}
Since it holds that $B^2+4 A^2=1$, one obtains that,
\begin{eqnarray}\label{chernint}
\left ( \frac{\partial \hat h}{\partial x_1}\times \frac{\partial \hat h}{\partial x_2}\right ) \cdot \hat h = -\frac{1}{y}\, k^2 B'.
\end{eqnarray}

In power of such expressions we determine the Chern-Simons index by integrating on a disk whose diameter is equal to the unit cell size, $R=L/2$,
\begin{eqnarray}\label{topo2}
Q= \frac{1}{4\pi}\int_0^{2\pi} d\theta \, \int_{0}^{k_0R} d\vert \vec x \vert \, \vert \vec x \vert \, \big (\frac{\partial \hat
h}{\partial x_1} \times \frac{\partial \hat h}{\partial x_2},
\big)\cdot \hat h
\end{eqnarray}
to obtain that,
\begin{eqnarray}
Q=-\frac{1}{2}\left [B(k_0R)-B(0) \right ],
\end{eqnarray}
since $B(0)=1$   the result of Eq.(\ref{chernvalue0}) is obtained.

\bibliography{biblio}
\end{document}